\documentclass[twocolumn]{aastex63} 

\newcommand{\kms}{$\rm km\;s^{-1}$}
\newcommand{\dg}{^\circ}

\def\degree{\mbox{$^{\circ}$}}

\def\kms{km\thinspace s$^{-1}$}

\def\eg{e.g.,~}
\def\etal{et al.~}

\begin{document}
\shorttitle{GALAXIES IN THE BO\"{O}TES VOID}
\shortauthors{Wegner \etal}


\title{METAL ABUNDANCES and STAR-FORMATION RATES OF EMISSION-LINE GALAXIES IN AND AROUND THE BO\"{O}TES VOID\\}

\author{Gary A. Wegner}
\affiliation{Department of Physics \& Astronomy, Dartmouth College, Hanover, NH, USA}

\author{John J. Salzer}
\affiliation{Department of Astronomy, Indiana University, Bloomington, IN, USA }

\author{Joanna M. Taylor}
\affiliation{Space Telescope Science Institute, Baltimore, MD, USA}

\author{Alec S. Hirschauer}
\affiliation{Department of Astronomy, Indiana University, Bloomington, IN, USA }
\affiliation{Space Telescope Science Institute, Baltimore, MD, USA}

\begin{abstract} We explore the possible dependencies of galaxy metal abundance and star-formation rate (SFR) on local environment, focusing on the volume of space in and around the Bo\"{o}tes Void.   Our sample of star-forming galaxies comes from the second catalog of the H$\alpha$-selected  KPNO International Spectroscopic Survey (KISS) which overlaps the void.  This sample represents a statistically complete, line-flux-limited ensemble of 820 star-forming galaxies, all of which possess metallicity and SFR estimates.   We carry out two distinct analyses of the KISS galaxies: one which probes the properties of the entire sample as a function of local density, and a second which details the properties of 33 KISS star-forming galaxies located within the Bo\"{o}tes Void.   In both cases we find no evidence that either the metallicity of the KISS galaxies or their SFRs depend on the environments within which the galaxies are located.    Our global analysis does show weak trends for decreasing stellar mass, decreasing metallicity, and decreasing SFRs with decreasing local densities.   However, we argue that the metallicity and SFR trends are artifacts of the stellar mass - local density trend.   In particular, the change in metallicity with density is precisely what one would predict from the mass-metallicity relation given the observed drop in stellar mass with decreasing metallicity.  Likewise, the SFR trend with density disappears when one instead considers the mass-normalized specific SFR.    The KISS galaxies dwelling in the Bo\"{o}tes Void are found to have nearly identical metallicity and SFR properties to a matched comparison sample, despite the fact that the former are located in density environments that are, on average, more than 16 times lower.
\end{abstract}


\section{Introduction}\label{sec:intro}

This paper presents an analysis of the chemical abundances and star-formation rates of a large sample (N = 820) of emission-line galaxies located in fields that cover the Bo\"{o}tes Void and its immediate surroundings.  

Voids in the large-scale structure of the universe can yield information on cosmological parameters, be used to study dark energy, and test non-standard models of gravity \citep[\eg][]{liebscher1992, moss2011, li2012, clampitt2013, lam2015}.  Substantial numbers of voids are now found from redshift data \citep[\eg][]{patiri2006, platen2007, pustilnik2011, pan2012, varela2012, pustilnik2013, mao2017, libeskind2018} using a variety of void-finding algorithms.  A large void is typically 20-50 Mpc in diameter with an under density $\delta_v$ = $\rho_v$/$\langle \rho \rangle$ - 1 $\sim -0.9$. Thus in addition to applications to cosmology, the low densities in voids relative to those in galaxy clusters and filaments provide different environmental conditions and an opportunity for studying the role of environment on galaxy formation and evolution.

The growth of voids has been studied by many authors using cosmological models \citep[for some recent reviews, see][]{vdw2011, einasto2011, cautun2014}.  Due to its underdensity, a void expands faster than the Hubble flow; matter flows from the void center as it expands and forms a higher density wall where the more rapidly expanding inner shells of material catch the slower expanding outer shells at the mature `shell crossing' phase. Recent  models find that structure inside voids is complex and voids will contain a weak filament structure \citep{aragcalvo2013, alpaslan2014}.  Simulations indicate that galaxies formed in lower-density regions have younger stellar ages, which appears to be consistent with observations \citep[\eg][]{delucia2006, kreckel2012}, but standard $\Lambda$CDM theory indicates that the underdense voids would contain many low mass dark matter halos.  These halos' galaxian counterparts are unobserved, however, which creates `the void phenomenon' along with the `too big to fail problem' which challenge the $\Lambda$CDM model \citep{peebles2001, hoeft2006, tikhonov2009, peebles2010, nasonova2011, papastergis2015}.   \citet{tinker2009} proposed a solution which requires there to be a critical mass threshold below which void galaxies cannot form enough stars and consequently are dark.  \citet{saintonge2008},  \citet{peebles2010},  \citet{kreckel2011c}, and  \citet{pan2012} point out, however, that such an arrangement may be somewhat contrived and that even if there are no observable stars, H I should be detectable. Nevertheless, these considerations indicate that void galaxies might differ from those in higher-density regions. The environmental conditions inside the voids and their walls could produce observable effects on the evolution and properties of the galaxies there, such as reduced star formation rates and lower chemical enrichments.  However, this simple expectation has not been found to be the case in most studies to date.

Many studies of void galaxies' properties, including optical spectra and photometry, have been published \citep[\eg][]{popescu1997, popescu1999, grogin2000, hogg2004, rojas2005, collobert2006, patiri2006, wegner2008, hoyle2012}.  Combined with H~I surveys \citep{szomoru1996, huchtneier1997, kreckel2011a, kreckel2012, kreckel2015, ricciardelli2014, moorman2014, moorman2015, moorman2016, douglas2018}, this has resulted in a general consensus that the lower density environments within which the void galaxies reside has only a small effect on their properties.   Rather, most authors conclude that the characteristics of galaxies in voids are more strongly set by the immediate surroundings associated with their dark matter halos.  

These previous studies have found that, generally, galaxies observed in voids are lower-mass systems with blue colors.   Void galaxies do not differ significantly  in structure from the same types of objects found in higher-density regions.   Some studies have found apparent differences in key properties for at least some void galaxies.  From a dynamical study of two early-type void galaxies, \citet{corsini2017} found that these objects possess lower dark-matter content than  early-type galaxies in high-density environments.  Furthermore, a handful of very metal poor and high $M/L$ void galaxies have been found \citep{kreckel2011b, pustilnik2013}, highlighting the importance of further searches for low-luminosity systems. 

The Bo\"{o}tes Void is one of the most studied voids in the nearby universe.  Following \citet{koss81, koss83, koss87} the void is located between roughly $\alpha = 206\dg - 240 \dg$ and $\delta = +32 \dg - +57 \dg$ in the redshift range $z = 0.04 - 0.06.$  Early work concentrated on optical spectroscopy and imaging \citep{tifft1986, strauss1988, weistrop1988, dey1990, peimbert1992, weistrop1992, weistrop1995, cruzen2002} and objective prism surveys \citep{case1982, case1987, moody1987}.   The results of these investigations support the conclusions described above:  void galaxies have properties consistent with their morphologies and luminosities, being typically gas rich blue disk galaxies (although some early-types are present).  These earlier studies provide a heterogeneous set of objects for looking at the properties of emission-line galaxies (ELGs) in the Bo\"{o}tes Void region.   We have therefore undertaken a new project to study the ELGs within and around the void.

The current study utilizes a large sample of ELGs cataloged as part of the KPNO International Spectroscopic Survey \citep[KISS;][]{salzer2000}, all of which possess self-consistent metallicity and star-formation rate (SFR) estimates.   A primary goal of this project is to explore possible environmental dependencies on these two parameters.  We carry out this study using a two-pronged approach.  The first is to look at the characteristics of the KISS ELG sample as a function of local environment.  This part of the sample utilizes the full catalog of KISS star-forming galaxies in the Bo\"{o}tes Void field.   The second portion of the study looks at the metallicities and SFRs of the specific KISS galaxies that are located within the Bo\"{o}tes Void and compares them with those found in the surrounding higher-density regions.  A standard $\Lambda$CDM cosmology with $\Omega_m = 0.3$, $\Omega_\Lambda = 0.7$, and $H_0 = 70$~kms$^{-1}$ Mpc$^{-1}$  is assumed in this paper.

\section{KISS Emission-Line Galaxy Sample}

\subsection{Description of Spectroscopy}

KISS employed a low-dispersion objective prism on the 0.61 m Burrell Schmidt telescope on Kitt Peak to carry out a comprehensive survey of ELGs in the nearby universe.   The objective-prism spectra covered two distinct wavelength ranges: $\lambda\lambda$ 6400-7200 (KISS red (KR), selected primarily by the H$\alpha$ line \citep{salzer2001, gronwall2004b, jangren2005a}) and $\lambda\lambda$ 4800-5500 (KISS blue, selected primarily by the [O III]$\lambda$5007 line; \citep{salzer2002}).  The second KISS red spectral survey strip \citep[hereafter KR2;][]{gronwall2004b} was centered on the declination of 43.5$\degree$ and traversed the entire width of the Bo\"otes Void slightly below its putative center.   The full RA coverage of KR2 was 11$^h$ 55$^m$ to 16$^h$ 15$^m$, and it had a declination extent of $\sim$1.8$^\circ$.   A total of 1029 ELG candidates were cataloged in KR2 in a survey area of 65.8 deg$^2$, including 12 that were identified as likely residing within the void based on their objective-prism redshifts.

A program of quick-look follow-up spectroscopy has been carried out by members of the KISS collaboration.   Spectroscopic observations for the KISS ELGs are presented in \citep{wegner2003, gronwall2004a, jangren2005b, salzer2005b}.  These spectra were necessary for determining accurate redshifts and for ascertaining the activity type of each galaxy (e.g., star forming {\it versus} AGN).   In addition, the spectra were used to infer metallicity estimates for the star-forming galaxies in KISS \citep[e.g.,][]{melbourne2002, melbourne2004, lee2004, salzer2005a, hirschauer2015, hirschauer2018}.   A large number of spectra of KISS galaxies located in and around the Bo\"otes Void were obtained specifically for this project using the MDM Observatory 2.4-m telescope and the Ohio State Boller and Chivens spectrograph.  

\begin{figure}[t]
\centering
\includegraphics[width=3.35in]{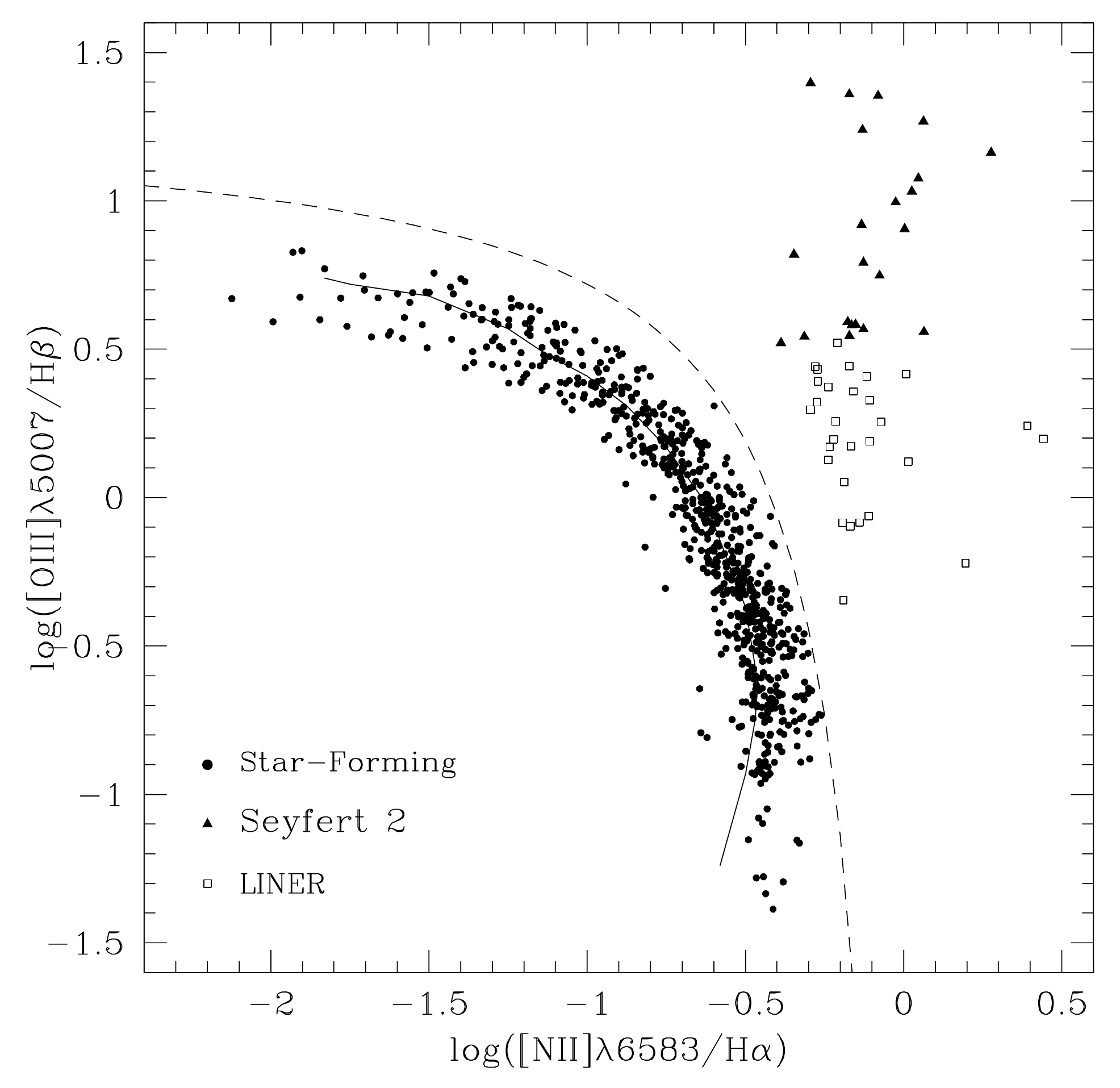}
\caption{Spectroscopic diagnostic plot for the full sample of KR2 emission-line galaxies.   Star-forming galaxies (SFGs) are plotted as dots, while Seyfert 2 galaxies and LINERs are plotted as triangles and squares, respectively.   The solid line that coincides with the SFGs represents a locus of model \ion{H}{2} regions from \citet{DE86}, while the dashed line is an empirical curve separating SFGs from AGN \citep{kauff03}.   Note that not all KR2 galaxies are plotted here, since in some cases the follow-up spectra are not of sufficient quality to yield reliable measurements of the four relevant emission lines (see text for details). \label{kr2bpt}}
\end{figure}

All 1029 KR2 galaxies possess follow-up spectra.   Of these, 101 were found to be false detections from the objective-prism survey, and 25 were detected via their [O III]$\lambda$5007 line \citep[see][]{salzer2009}, and hence are at redshifts well beyond the volume of space under consideration in this study.   The remaining 903 galaxies are {\it bona fide} ELGs located within the redshift range covered by the KISS red survey (0 to 0.095).    Figure~\ref{kr2bpt} shows an emission-line ratio diagnostic diagram (a.k.a.  BPT diagram; \citet{bpt, vo1987}) of the KR2 galaxy sample.  The location of each object in Figure~\ref{kr2bpt}  allows them to be classified into their appropriate activity type.  Based on the information present in their spectra, we classify 820 as star-forming ELGs, while the remaining 83 are AGN (either Seyfert 1 (9), Seyfert 2 (23) or Low-Ionization Nuclear Emission Region (LINER, 51)).  

\subsection{Properties of the Sample}

The current study focuses on the metallicity and SFRs of the KISS star-forming galaxies as a function of their local environment.   Metallicities are derived using the O3N2 abundance calibration presented in \citet{hirschauer2018}.   For 70 KR2 galaxies the existing follow-up spectra do not contain the necessary emission lines to allow us to compute an O3N2 abundance.   For these galaxies the abundances are estimated using the mass-metallicity (M-Z) relation presented in the \citet{hirschauer2018} study.   SFRs are calculated using the \citet{kennicutt1998} relation: SFR =  L$_{H\alpha}$/7.9 $\times$ 10$^{42}$, where the H$\alpha$ luminosities are derived from the distances to each KISS galaxy inferred from their redshifts, along with the H$\alpha$ fluxes measured from the discovery objective-prism spectra.   The latter were considered to be a superior measurement of the {\it total} H$\alpha$ emission for each KISS galaxy, as opposed to using the fluxes from the slit spectra.

\begin{figure}
\centering
\includegraphics[width=3.35in]{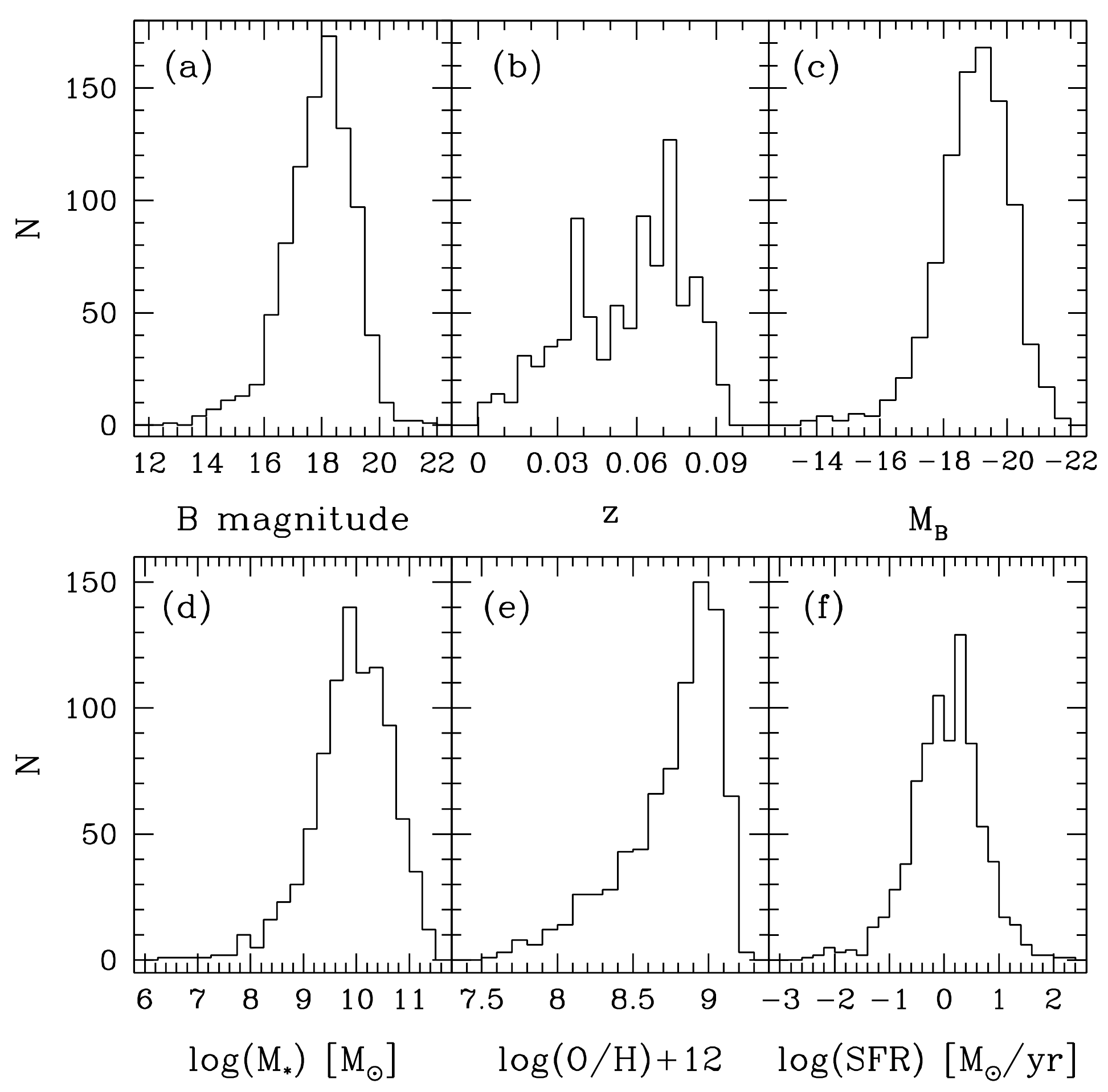}
\caption{Histograms exhibiting the basic characteristics of the KR2 sample.  Panels (a) - (d) include all 903 KR2 galaxies that are H$\alpha$-detected, while panels (e) - (f) plot only the star-forming ELGs (N=820).  (a) Apparent B magnitudes for the KR2 galaxies; median B = 18.02.  (b) Measured redshifts; median z = 0.0615.  The survey has a hard upper redshift limit of 0.095.  (c) Absolute B-band magnitudes; median M$_B$ = $-$19.05.  (d) Log(M$_\star$), the stellar mass in M$_\odot$; median log(M$_\star$) = 9.96 M$_\odot$.  (e) Metallicity values for the KR2 star-forming galaxies; median log(O/H)+12 = 8.86.  (f) Log(SFR), the star-formation rate in M$_\odot$/yr; median log(SFR) = 0.07 M$_\odot$/yr. \label{kr2hist}}
\end{figure}

A series of histograms illustrating the basic properties of KR2 galaxies is shown in Figure~\ref{kr2hist}.  The apparent B-band magnitude distribution (panel (a)) has a median value of B = 18.02, which is comparable to the magnitude limit of the Sloan Digital Sky Survey (SDSS) redshift survey \citep[\eg][]{strauss2002}.   Only 44\% of the KR galaxies were observed spectroscopically by SDSS  \citep{hirschauer2018}, which means that at any distance in the volume under study the KISS galaxies reach between 1 and 2 magnitudes deeper than the SDSS redshift survey (see below).  The absolute magnitudes plotted in panel (c) have a median value of M$_B$ = $-$19.05, roughly one magnitude fainter than M$^*_B$, the characteristic luminosity that represents the location of the ``knee" in the \citet{schechter1976} luminosity function.  For comparison, the luminosity of the LMC is roughly 0.5 magnitudes fainter than the KR2 galaxy median value \citep{mcconnachie2012}.  The majority of the KR2 galaxies possess intermediate luminosities, although $\sim$10\% have luminosities fainter than the SMC (M$_B$ = $-$17.5), making them true dwarf galaxies.   Stellar masses (panel (d)) were derived using a multi-wavelength SED fitting method (UV through IR; see \citet{hirschauer2018} for details).  The panels showing the O/H abundances and SFRs (panels (e) and (f)) plot only the N = 820 star-forming KR2 galaxies.   The distribution of derived abundances is peaked around the median value of log(O/H)+12 = 8.86, but with a strong tail to lower values.   We note that the KISS abundances are nearly always a measurement of the central value, which tend to be systematically higher by up to 0.5 dex compared to abundances measured further out in the disk.  The median SFR is $\sim$1 M$_\odot$/yr, consistent with expectations for a sample of intermediate luminosity star-forming galaxies.  A modest tail reaches up to SFR $\sim$100 M$_\odot$/yr, while a handful of the dwarf KR2 galaxies have SFR $<$0.01 M$_\odot$/yr.

\section{Density Analysis of KR2 Volume}\label{sec:dens}

In this section we describe our methodology for characterizing the density environment within the volume of space covered by the KR2 ELG survey.  Since a primary goal of this project is to study the effects of environment on the abundances and star-formation rates of the KR2 galaxies, it is important that we do not utilize the KISS sample to characterize the environment.   Rather, we use an independent, comprehensive redshift catalog of galaxies to describe the density distribution.

We utilize the galaxy redshifts available from the Sloan Digital Sky Survey (SDSS; Data Release 7 \citep{sdss7} was used), supplemented with redshifts for the brighter galaxies in our survey area taken from the Updated Zwicky Catalog (UZC; \citet{falco1999}) to define the distribution of galaxies in our survey volume.   We searched these two catalogs for all objects located in an extended region that includes the KR2 survey area as well as a generously-sized bordering region.   The region covered by our comparison catalog is 11$^h$ 44$^m$ to 16$^h$ 28$^m$ (176$^\circ$ -- 247$^\circ$) in RA and 40$^\circ$ to 47$^\circ$ in Dec.   This allows for a border around the KR2 survey region of at least 2.5$^\circ$  on each side.  This border region allows us to derive accurate densities for the galaxies found in KR2 without the impact of any serious edge effects (see below).   Figure~\ref{skymap} shows the geometry of our survey regions, with the smaller KR2 survey area denoted by the red rectangle.  The total number of objects in the comparison sample of SDSS+UZC catalog galaxies is 14577. 

A cone diagram illustrating the distribution of the galaxies in our survey volume is shown in Figure~\ref{cone30}.  The redshift range covered by the cone is limited to velocities under 30,000 \kms.   This is because the KISS red survey is limited to galaxies with z $<$ 0.095 ($\sim$28,500 \kms) due to the filter used to obtain the objective-prism images \citep{salzer2000}.   The KISSR galaxies are plotted as small red circles, while the comparison sample of SDSS+UZC galaxies are shown as small black dots.  The SDSS+UZC galaxies shown cover the full seven degree declination range of the comparison sample, while the KR2 galaxies are only found in the central 1.8$^\circ$ region denoted in Figure~\ref{skymap}.  The large-scale structure within the survey volume is very well delimited by the galaxies in the comparison sample.   The large red circle in Figure~\ref{cone30} denotes the location of the Bo\"otes Void  (see Section~\ref{sec:void}).   

\onecolumngrid

\begin{center}
\begin{figure}
\includegraphics[width=6.5in]{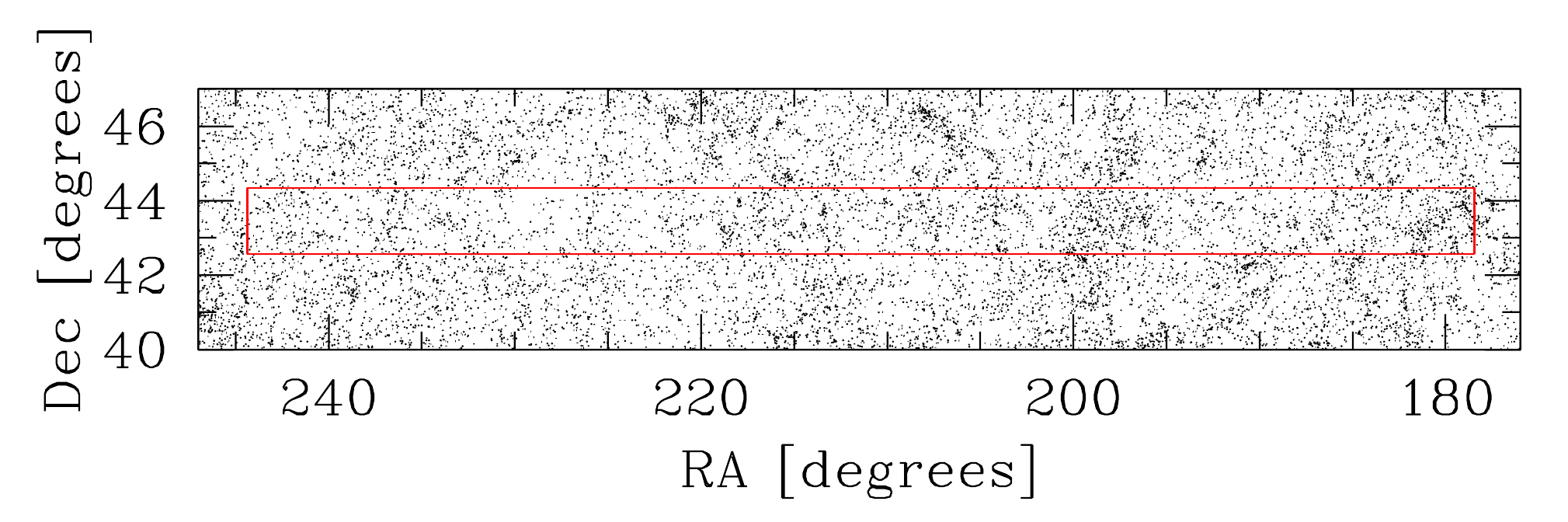}
\caption{Sky map showing the region considered in this study.   The red rectangle denotes the area covered by the KR2 survey (RA range 11$^h$ 55$^m$ to 16$^h$ 15$^m$ (178.5$^\circ$ -- 244.4$^\circ$), Dec range 42.55$^\circ$ -- 44.35$^\circ$).  The comparison sample of galaxies with redshifts in the UZC and SDSS catalogs covers the RA range of 11$^h$ 44$^m$ to 16$^h$ 28$^m$ (176.0$^\circ$ -- 247.0$^\circ$) and 40.0$^\circ$ to 47.0$^\circ$ in Dec.  Each small black dot represents a galaxy in the comparison catalog (N=14577).  The lower density of dots in the RA range $\sim$220$^\circ$ -- 230.0$^\circ$ coincides with the central region of the Bo\"{o}tes Void. \label{skymap}}

\includegraphics[width=6.5in]{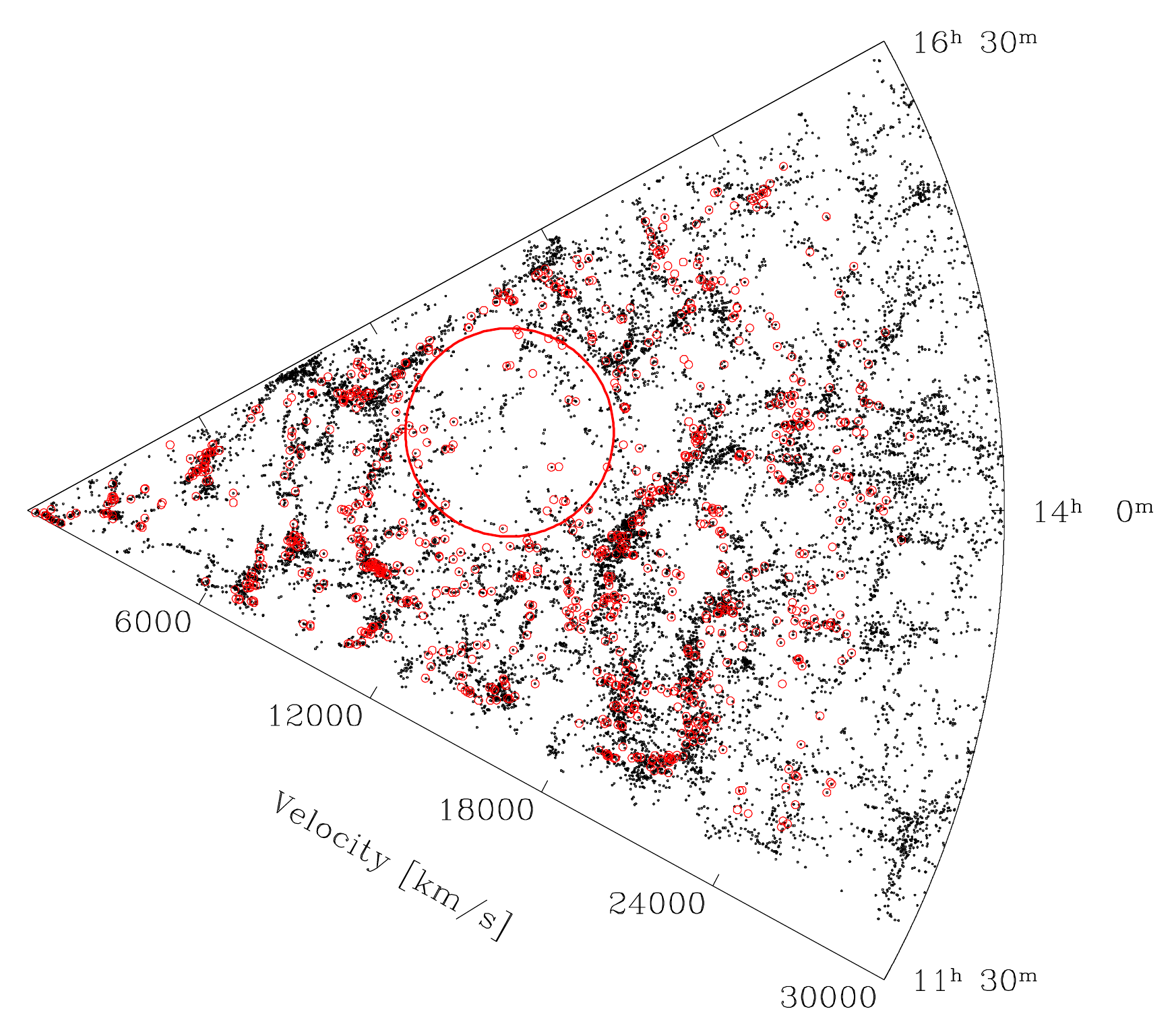}
\caption{Cone diagram for the extended region considered in this study.  This covers the full volume of the comparison sample (7$^\circ$ in declination) shown in Figure~\ref{skymap} out to a redshift of 30,000 \kms.   The KISS survey covers only the middle $\sim$1.8 degrees of this volume (Declination extent).   Black dots are comparison galaxies, red circles are KISS galaxies.   The large red circle denotes the location of the Bo\"{o}tes Void used in this study. \label{cone30}}
\end{figure}
\end{center}
\twocolumngrid



We assess the local density environment using a standard N-nearest neighbors algorithm.  Our analysis excludes all galaxies with observed velocities less than 2100 \kms (z $<$ 0.007) since for these galaxies redshift is not a reliable indicator of distance.   Furthermore, for these small distances the survey boundaries are all very close to the galaxies, making the determination of the local density less accurate.  There are only ten KR2 galaxies with z $<$ 0.007, meaning that the final sample used in the density analysis is N = 893.  For each galaxy in the input catalog, the 3D separation is computed to all galaxies in the SDSS+UZC comparison catalog.   The distances to the N nearest galaxies are tracked until the entire comparison catalog has been searched.  The density in the vicinity of the target galaxy is then computed as N divided by the spherical volume out to the distance of the Nth nearest neighbor.   We experimented with the value of N that gave the most robust results, trying values between 5 and 20.   We settled on N = 15 for the current study, since it provides a good balance between the desire to make N large enough to provide adequate sampling while not making it so large that survey edge effects became important.

\begin{figure}
\centering
\includegraphics[width=3.20in]{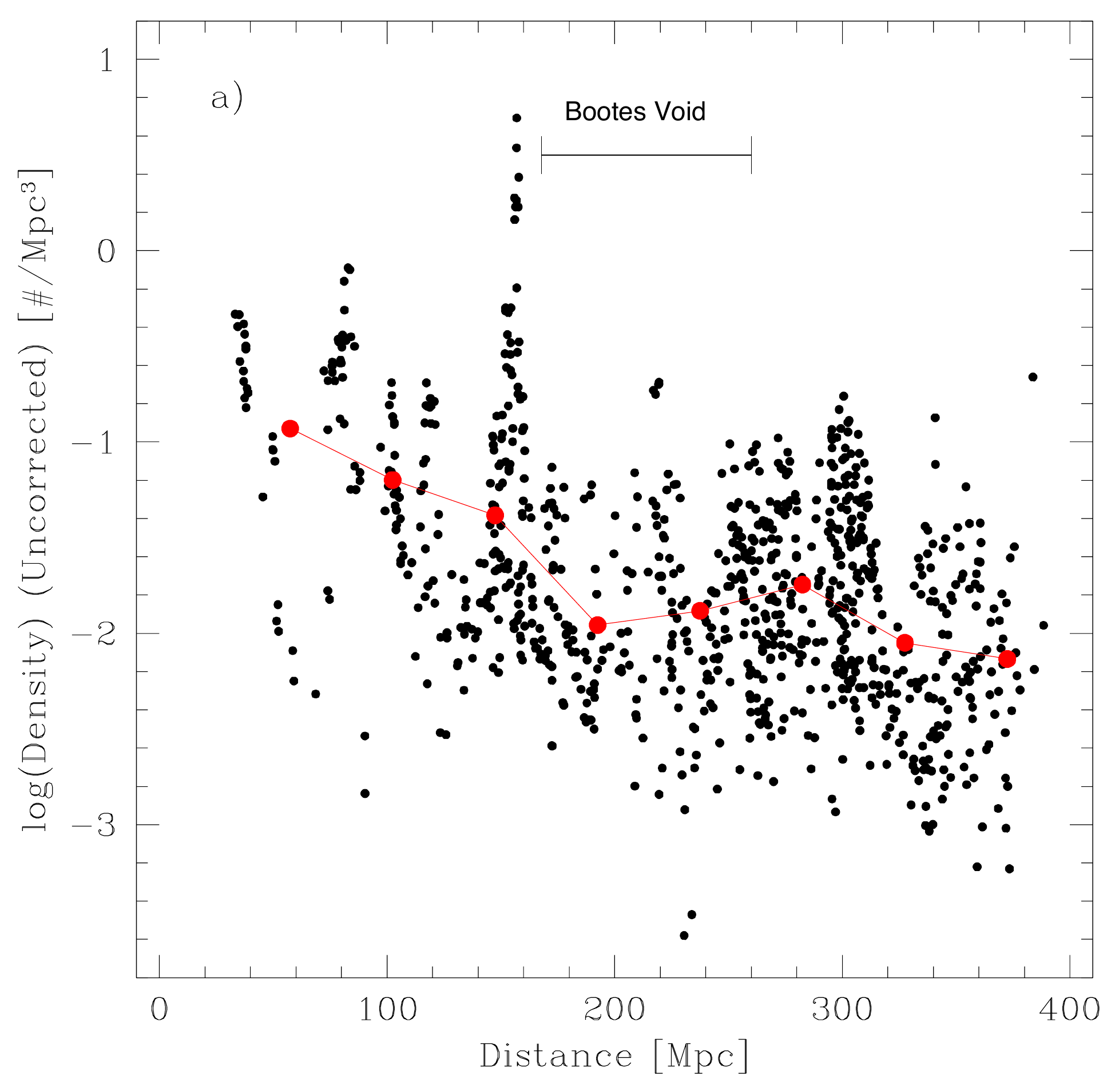}
\includegraphics[width=3.20in]{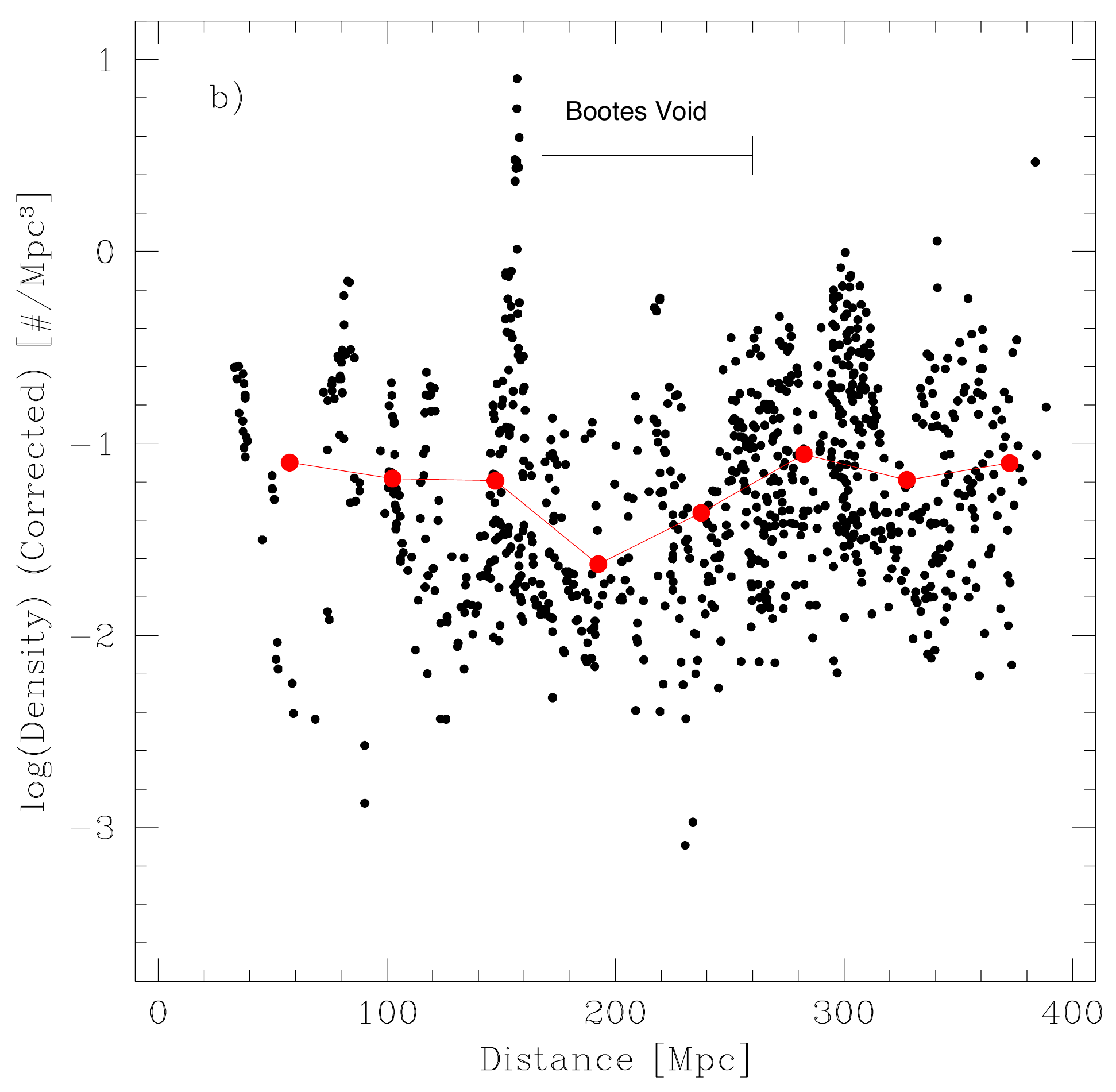}
\caption{Uncorrected (a) and corrected (b) volume densities for the KR2 galaxies.  The red points in both plots represent the mean value for log(density) in evenly spaced (45 Mpc separation) distance bins.   The drop on the mean density at the distances covered by the Bo\"{o}tes Void is evident in both plots, despite the fact that the averages are computed for {\it all} galaxies at these distances, not just those in the RA range covered by the void.  In the corrected data the mean density is roughly constant outside the region occupied by the Bo\"{o}tes Void, indicating the validity of the corrections applied. \label{denscorr}}
\end{figure}

As alluded to above, our use of the extended border region surrounding the KR2 survey volume (Figure~\ref{skymap}) meant that we rarely encountered this edge-of-survey problem.    For example, when the densities for the KR2 galaxies were derived the 15th galaxy was encountered prior to the search radius reaching the nearest edge of the survey volume in 94.2\% of the cases.  For the handful of KR2 galaxies that were affected by the edge-of-survey limitation, we resorted to computing the density using those galaxies that were located within the distance associated with the nearest survey boundary.   In 98.3\% of the cases, the density was computed using at least 5 nearest neighbors.   All cases where the survey edge was reached prior to the 15th galaxy being found fall at lower redshifts, closer to the apex of the wedge-shaped volume.  At the distances associated with the Bo\"otes Void the linear dimensions of the border regions were sufficient to ensure that edge effects were not an issue.

Since the SDSS+UZC comparison catalog is magnitude limited, it suffers from increasing incompleteness at larger distances.  Hence, the densities derived using this catalog will decrease as the distance increases.  This is seen in the left panel of Figure~\ref{denscorr}, which plots the uncorrected volume densities derived for the galaxies in the KR2 sample.   There is a strong distance dependence on the mean density (red points). This effect is easily compensated for using the luminosity-function normalization method \citep{postman1984}, which corrects the observed number densities for decreasing sampling with redshift.   We have corrected our densities using this method, employing the luminosity function parameters derived by \citet{blanton2003} for the SDSS galaxies (M$_g^*$ = $-$20.1 and $\alpha$ = $-$0.89).  The results of this 

\onecolumngrid

\begin{center}
\begin{figure}[t!]
\centering
\vskip 0.5in
\includegraphics[width=7.0in]{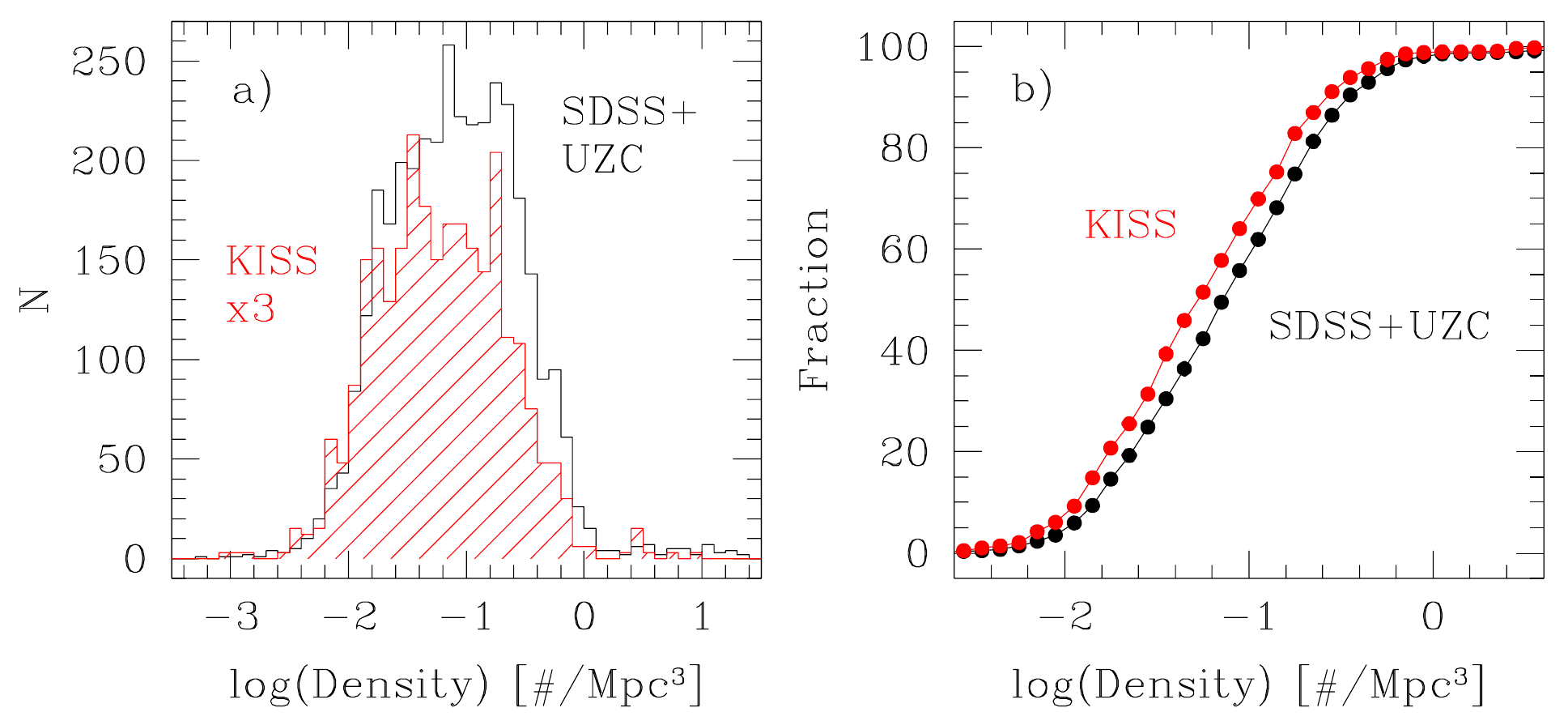}
\caption{(a) Histograms showing the distributions of densities computed for the KR2 galaxies (red; N=893) and for the comparison SDSS+UZC sample galaxies located within the celestial footprint of the KR2 survey volume (black, N=3548).  The KR2 histogram has been scaled up by a factor of three for better comparison with the SDSS+UZC sample.  The distributions of densities appear similar, although the KISS sample is offset slightly to lower densities   (b) Cumulative density distribution plots for the same samples plotted in panel (a).   This plot clearly shows the modest offset in density between the two samples, in the sense that the KISS galaxies are located at slightly lower densities on average.  The offset in the median densities is 0.14 dex galaxies/Mpc$^{-3}$, meaning that the KISS galaxies have a median density 27\% lower than the SDSS+UZC sample. \label{denshist}}
\end{figure}
\end{center}
\twocolumngrid

\noindent correction are shown in the right panel of Figure~\ref{denscorr}, where the median density values in representative distance bins are now constant outside the region covered by the Bo\"otes Void.

\begin{center}
\begin{figure}[t!]
\centering
\includegraphics[width=4.8in]{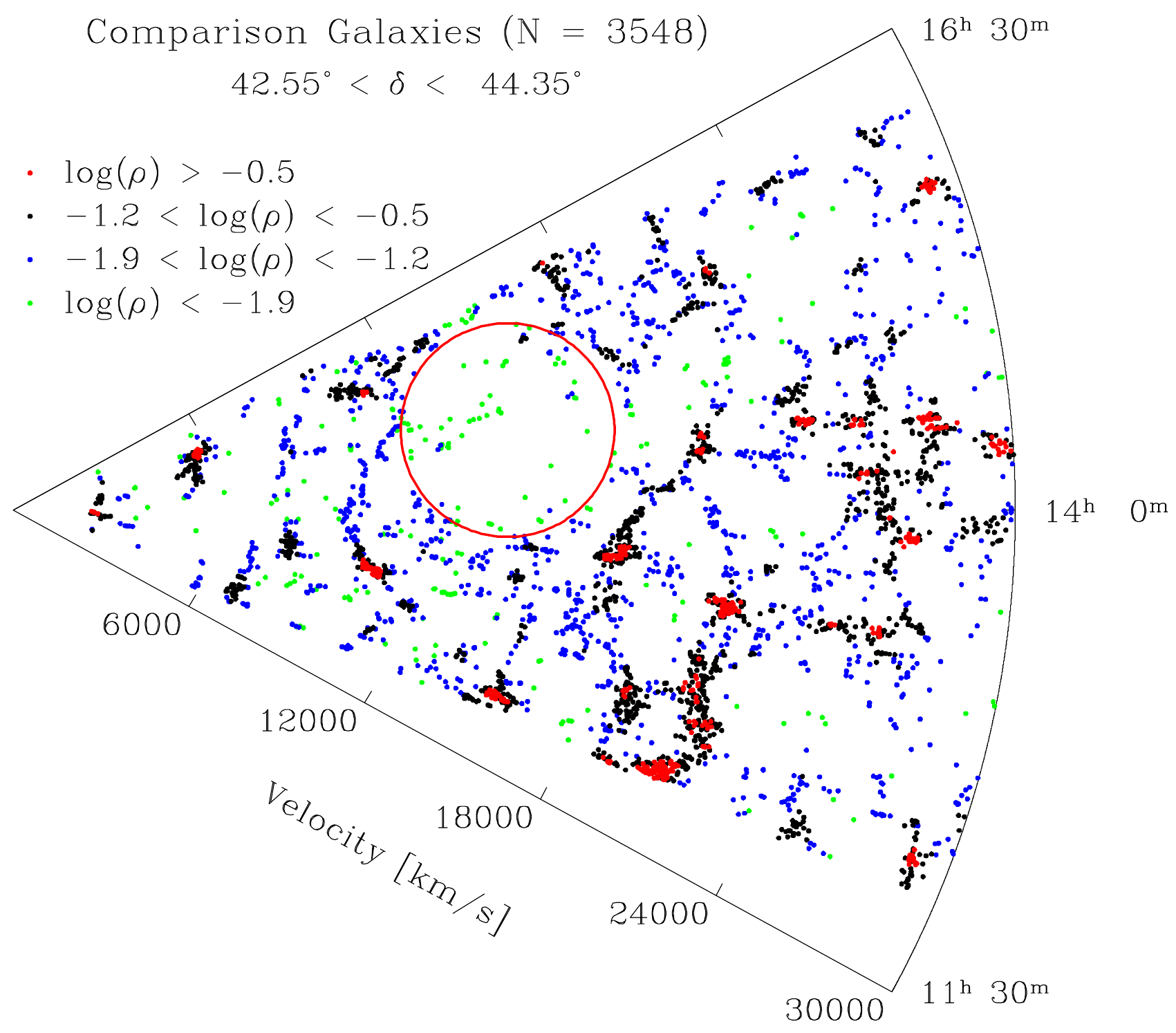}
\includegraphics[width=4.8in]{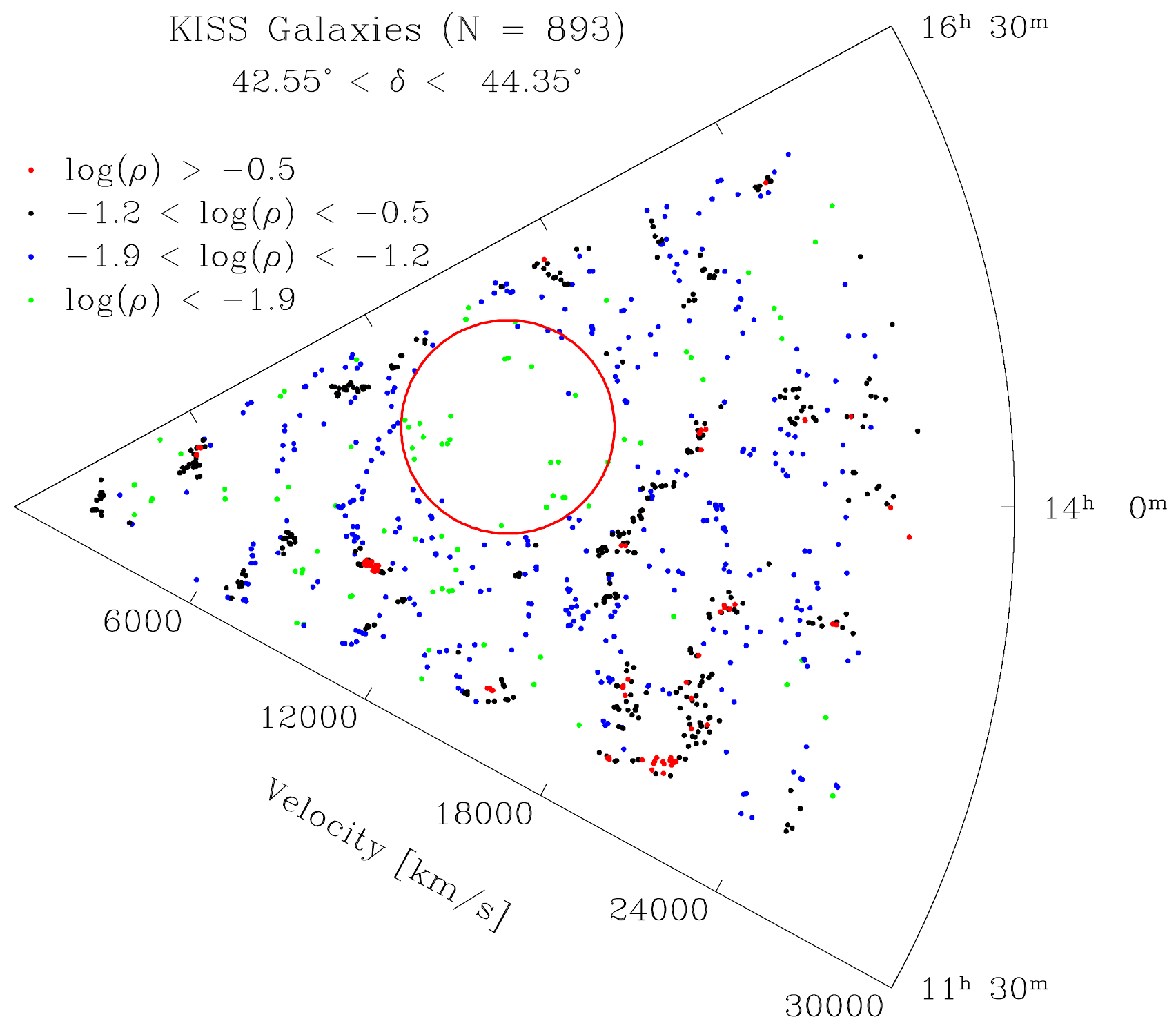}
\caption{Cone diagrams of the comparison and KISS galaxies located within the KR2 survey region (1.8$^\circ$ in declination), where the color of each point indicates the local galaxy density.   The upper plot shows the galaxies in the comparison SDSS+UZC sample that lie within the boundaries of the KR2 survey area (N = 3548).   Four density regimes are illustrated: high density (red points, log($\rho_{gal}$) $>$  $-$0.5 Mpc$^{-3}$), moderate density (black points, $-$0.5 $>$ log($\rho_{gal}$) $>$  $-$1.2), intermediate density (blue points, $-$1.2 $>$ log($\rho_{gal}$) $>$  $-$1.9), and low density (green points, log($\rho_{gal}$) $<$  $-$1.9 Mpc$^{-3}$).    The lower plot shows the locations of the KISS ELGs in precisely the same volume of space.   The color scheme used for the dots is identical to that of the upper figure.   The large red circle in both plots denotes the location of the Bo\"{o}tes Void used in this study.  
\label{denscone}}
\end{figure}
\end{center}

The results of our density computations are shown in Figures~\ref{denshist} and \ref{denscone}.  Figure~\ref{denshist}a presents histograms showing the distribution of galaxian densities for both the galaxies in the SDSS+UZC comparison catalog and the KR2 ELG sample.  The densities for the comparison sample were derived by first creating a sub-sample of SDSS+UZC galaxies that occupied the same volume of space as the KR2 galaxies.  For each comparison sample galaxy located within this volume, its density was derived using the full SDSS+UZC catalog in a manner exactly analogous to how the KR2 densities were computed.  Hence, the densities for the comparison galaxies and KISS ELGs are directly comparable.  The histograms show that the two samples overlap completely in terms of their density environments.   However, the two distributions are offset slightly, in the sense that the KISS galaxies are located in slightly lower density environments, on average.   The median densities are 0.0803 galaxies/Mpc$^3$ for the SDSS+UZC galaxies and 0.0583 galaxies/Mpc$^3$ for the KISS sample (27\% lower).   The median density of the star-forming sub-sample of the KR2 galaxies is only slightly lower:  0.0570 galaxies/Mpc$^3$.  Figure~\ref{denshist}b plots the same information but in the form of a cumulative distribution function (CDF).   A Kolmogorov-Smirnov (K-S) test applied to the CDFs confirms that the differences in the two distributions are strongly significant.   The probability that the two density distributions shown in Figure~\ref{denshist}a are drawn from the same parent population is essentially zero.

Cone diagrams that illustrate the spatial distribution of the density environments of the comparison and KISS ELG galaxies are shown in Figures~\ref{denscone}.    The local galaxian density is coded by the color of the symbols, binned into four density regimes: red for highest density, black  for moderately high, blue for intermediate, and green for low density (see Figure caption for precise density values for each group).   Consistent with the results shown in 
\onecolumngrid

\twocolumngrid

\noindent Figure~\ref{denshist}, the KR2 ELGs are found in all density regimes from high to low.   However, there is a higher percentage of KR2 galaxies in the intermediate and low density bins (51.5\%) compared to the comparison sample (42.3\%) again suggesting that the ELGs have a tendency to be located in slightly lower density environments than the SDSS+UZC galaxies.   

The results of the density analysis presented in this section are used in the subsequent sections to explore how the properties of the KISS ELGs vary as a function of local galaxy density.

\section{Global Characteristics of the KISS ELGs as a Function of Local Density}\label{sec:global}

As described in \S~\ref{sec:intro}, we divide our investigation into the properties of the KISSR galaxies as a function of environment into two components.   In the current section, we look at the physical characteristics of the full KR2 sample {\it versus} the local density environment.  This allows us to take advantage of the full catalog of KR2 ELGs.  In the subsequent section we focus on the properties of the KISSR galaxies located within the Bo\"otes Void.

We explore the characteristics of several physical properties as a function of local galaxy density: stellar mass (M$_*$), B-band absolute magnitude (M$_B$), oxygen abundance (log(O/H)+12), star-formation rate (SFR), and specific star-formation rate (sSFR = SFR / M$_*$).  In each case, we plot the given property for the relevant KR2 galaxy sample {\it versus} density.  Each diagram also shows the trend of the given property {\it versus} density for the KR2 star-forming galaxies by plotting median values computed using two different binning schemes.  In one case, the data are binned using a fixed bin width ($\Delta$(log($\rho$) = 0.25 dex; red dots in Figures~\ref{densplot_mass} -- \ref{densplot_ssfr}), while in the second the bin widths are variable but the number of galaxies in each bin is fixed as N = 64 (green dots in Figures~\ref{densplot_mass} -- \ref{densplot_ssfr}).

\subsection{Stellar Mass and Luminosity as a Function of Environment}

\begin{figure}
\centering
\includegraphics[width=3.35in]{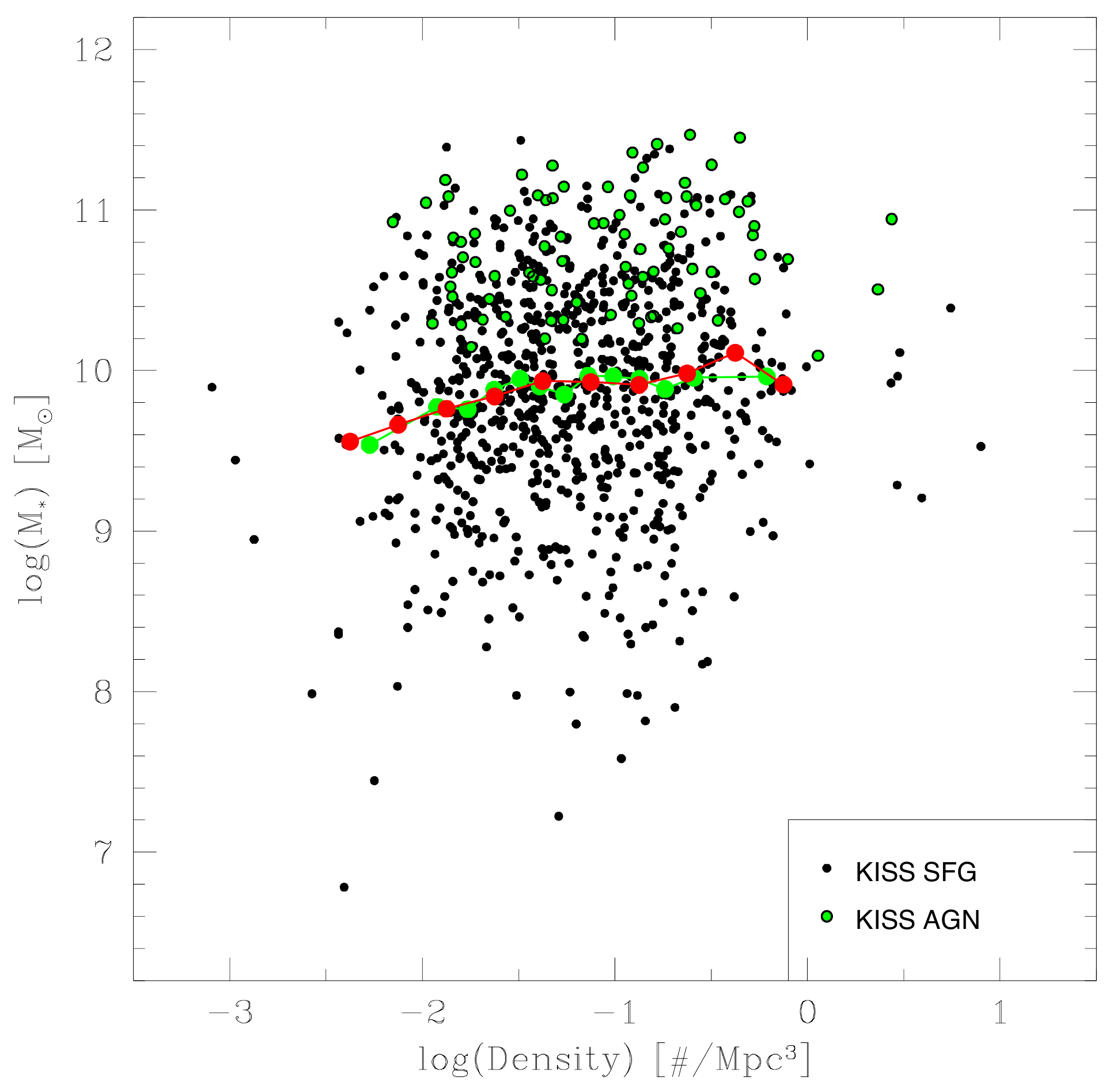}
\caption{Stellar mass (M$_*$) plotted versus local density (\# Mpc$^{-3}$) for the KISS galaxies.  The total number of galaxies plotted is N=893.  Of these, 83 are AGN (Seyfert 1's (N=9), Seyfert 2's (N=23) and LINERs (N=51)); the AGN are plotted as green dots.  The remaining 810 objects (black dots) are star-forming galaxies (SFGs).  The larger red and green dots represent the median stellar masses in binned ranges of density for the KISS SFGs only.  The green dots are binned such that there is a fixed number of galaxies in each bin (N=64); the spacing between the bins varies.  The red dots represent bins of fixed width (0.25 dex) but with variable numbers of objects per bin.  Only bins with 12 or more galaxies are plotted.   There is a trend for the median stellar mass to decrease with decreasing local density. \label{densplot_mass}}
\end{figure}

We begin by considering the density distribution of two parameters that quantify the size/bulk of the KR2 galaxies: stellar mass (M$_*$) and B-band absolute magnitude (M$_B$).   Figure~\ref{densplot_mass} plots the logarithm of stellar mass {\it versus} the logarithm of the local density for all 893 KR2 galaxies with density measurements.  This includes 83 KISS AGN, which appear in the figure as the smaller green dots.  Not surprisingly, the AGNs are strongly biased toward the higher masses: they have a median log(M$_*$) of 10.76 M$_\odot$, compared to 9.88  M$_\odot$ for the KISS SFGs (a factor of 7.6 times higher mass).   However, there is only a slight tendency for the AGNs to be located in higher density regions: the median local density for the AGN subsample is only a factor of 1.67 times higher than it is for the SFGs.  

\begin{figure}
\centering
\includegraphics[width=3.35in]{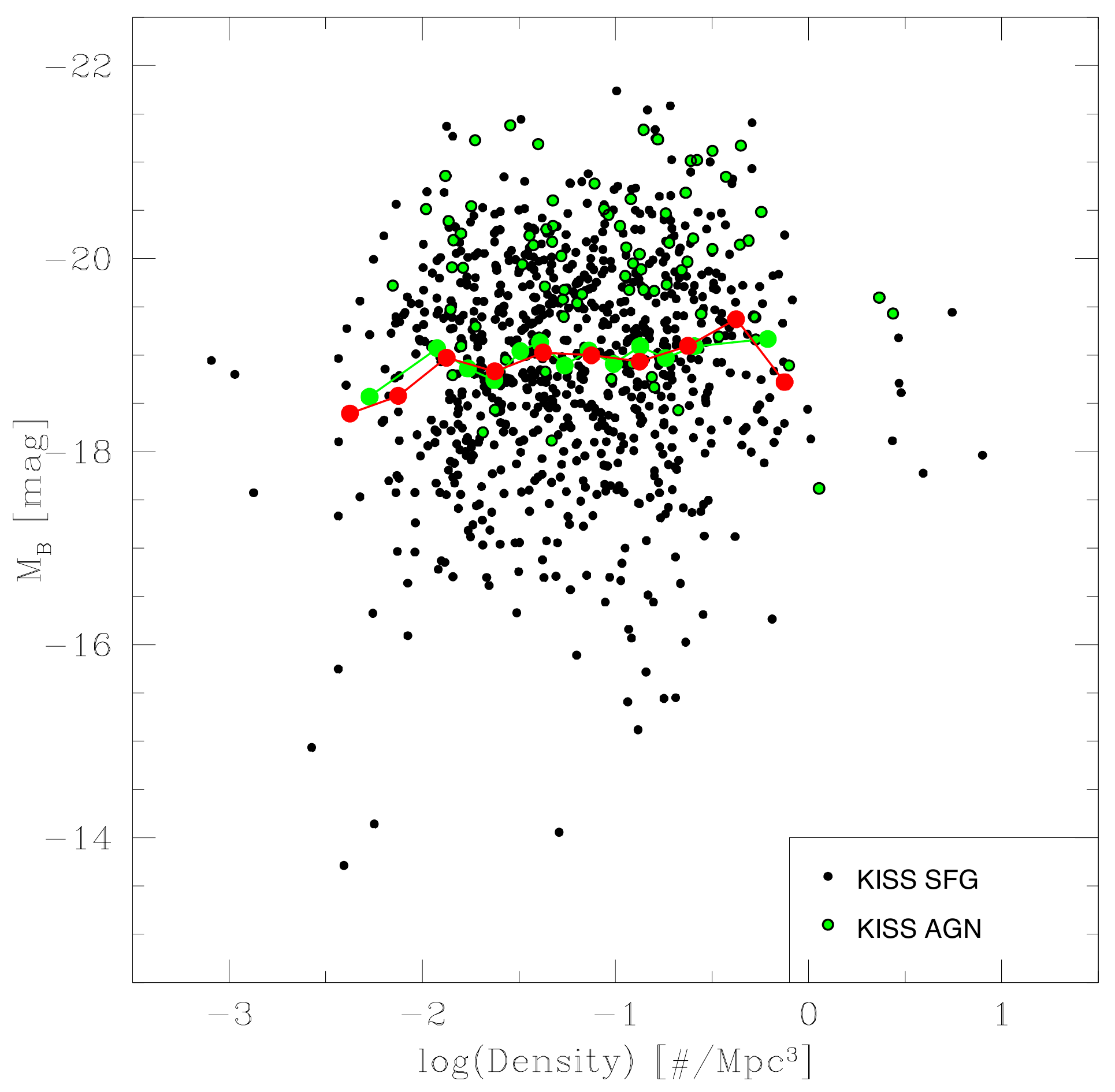}
\caption{B-band absolute magnitude (M$_B$) plotted versus local density (\# Mpc$^{-3}$) for the KISS galaxies.  The total number of galaxies plotted is N=893.  Of these, 83 are AGN (Seyfert 1's (N=9), Seyfert 2's (N=23) and LINERs (N=51)); the AGN are plotted as green dots.  The remaining 810 objects (black dots) are star-forming galaxies (SFGs).   The larger red and green dots represent the median absolute magnitudes in binned ranges of density for the KISS SFGs only.  The green dots are binned such that there is fixed number of galaxies in each bin (N=64); the spacing between the bins varies.  The red dots represent bins of fixed width (0.25 dex) but with variable numbers of objects per bin.  Only bins with 12 or more galaxies are plotted.  \label{densplot_lum}}
\end{figure}

Focusing now on the SFGs, it is clear that the KR2 galaxies are found with a full range of masses  at all density regimes populated by the sample; Figure~\ref{densplot_mass}  is essentially a scatter diagram.  However, the median values of the binned data do show a weak trend for the KR2 SFGs to  have lower masses at lower densities.  For densities above log($\rho$) $\sim$ $-$1.5 galaxies/Mpc$^3$ the median density is essentially constant (large red and green dots).   At lower densities the median stellar mass drops by 0.38 dex over 1.0 dex in density (a factor of 2.4).  

Similarly, Figure~\ref{densplot_lum} shows the distribution of B-band absolute magnitude {\it versus} local density for the KISS galaxies.   As in the previous figure, the KISS AGNs are plotted as small green dots, which trend toward the higher luminosities.  The median absolute magnitude of the KISS AGNs is nearly a full magnitude more luminous than the the correseponding value for the KISS SFGs ($-$19.91 for the AGNs {\it versus} $-$18.96 for the SFGs).   The KISS SFGs exhibit a fairly constant median absolute magnitude (large red and green dots) over most of the density range covered by our sample, then drop by nearly 0.6 mag over the two lowest luminosity bins (red points).   This roughly mimics the behavior of the stellar masses in Figure~\ref{densplot_mass}.   These results are consistent with previous studies that have found subtle differences in the clustering characteristics of lower luminosity/mass galaxies \citep[\eg][]{salzer1990}, in the sense that galaxies in voids tend to be lower mass systems.

\subsection{Metallicity and SFR as a Function of Environment}

Next we turn our attention to the main focus of this study: the role that local environment has on the metal abundance and star-formation characteristics of galaxies.  As stressed earlier, every star-forming galaxy in the KR2 sample possesses both a metallicity estimate and a measured SFR.   The complete nature of our sample means that our results are free from any selection biases beyond those inherent in the construction of the KISS ELG sample itself.

\begin{figure}
\centering
\includegraphics[width=3.35in]{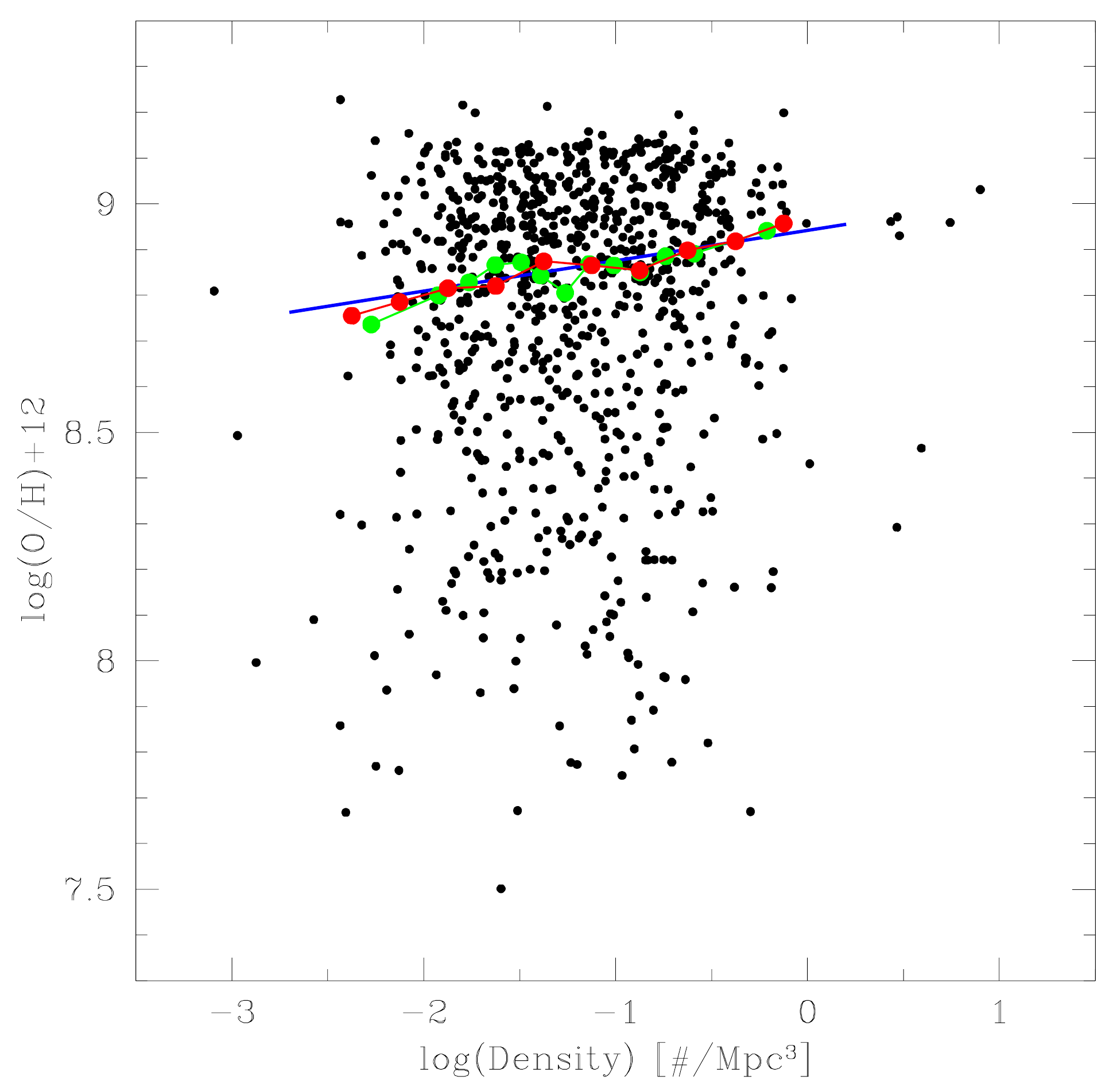}
\caption{Oxygen abundance plotted versus local density (\# Mpc$^{-3}$) for the KISS star-forming galaxies.  The total number of galaxies plotted is N=810.  These are the same objects as the black dots in Figure \ref{densplot_mass}.  The larger red and green dots represent the median oxygen abundances in binned ranges of density.  The green dots are binned such that there is fixed number of galaxies in each bin (N=64); the spacing between the bins varies.  The red dots represent bins of fixed width (0.25 dex) but with variable numbers of objects per bin.  Only bins with 12 or more galaxies are plotted.  Both sets of binned data show a trend for decreasing abundance as a function of local density.   For the fixed bin size data (red points) the decrease in the median abundance is 0.20 dex over a 2.5 dex drop in density.   The solid blue line represents a weighted linear fit to the red points, as described in the text. \label{densplot_abun}}
\end{figure}

Figure~\ref{densplot_abun} displays the distribution of measured oxygen abundances {\it versus} local galaxy density.  As with our previous plots, the individual KR2 galaxies populate a wide range of parameter space in both quantities being plotted.   The trend for the KR2 ELGs to possess high abundances is evident in the higher density of points near the top of the figure.  The median abundances in the density bins show a clear trend of decreasing abundance with lower galaxian density.   For example, the median abundance drops from 8.96 to 8.76 ($\Delta$log(O/H) = $-$0.20) over a 2.5 dex drop in local density (red points) between log($\rho_{gal}$) = 0 and log($\rho_{gal}$) = $-$2.5.   A weighted linear fit to the red points in Figure~\ref{densplot_abun} (solid blue line) yields a slope of 0.066 $\pm$ 0.022.   This fit results in a value of $\Delta$log(O/H) = $-$0.166 $\pm$ 0.069 between log($\rho_{gal}$) = 0 and $-$2.5.   Hence, while the data support a systematic drop in the metallicity of the KR2 galaxies at lower galaxian densities, the overall trend is small and of only moderate statistical significance (2.4 $\sigma$).

\begin{figure}
\centering
\includegraphics[width=3.35in]{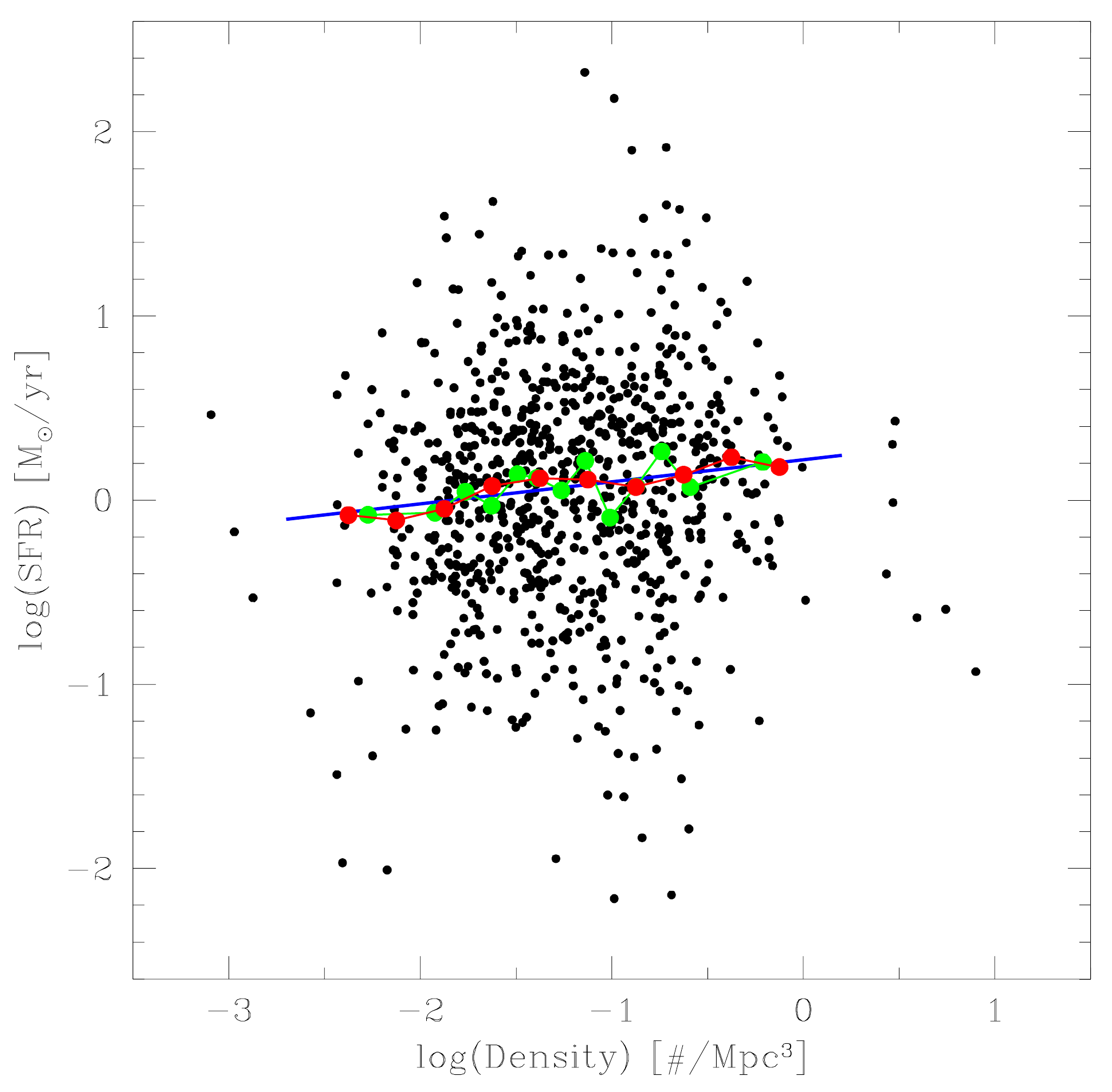}
\caption{H$\alpha$ star-formation rate (SFR) plotted versus local density (\# Mpc$^{-3}$) for the KISS star-forming galaxies.  The total number of galaxies plotted is N=810.  These are the same objects as the black dots in Figure \ref{densplot_mass}.  The larger red and green dots represent the median SFRs in binned ranges of density.  The green dots are binned such that there is fixed number of galaxies in each bin (N=64); the spacing between the bins varies.  The red dots represent bins of fixed width (0.25 dex) but with variable numbers of objects per bin.  Only bins with 12 or more galaxies are plotted.  Both sets of binned data show a weak trend for decreasing SFR as a function of local density.   For the fixed bin size data (red points) the decrease in median SFR is $\sim$0.26 dex over a 2.5 dex drop in density.  The solid blue line represents a weighted linear fit to the red points, as described in the text. \label{densplot_sfr}}
\end{figure}

The KR2 ELG sample exhibits a similar behavior in the plot showing the distribution of log(SFR) {\it versus} local density (Figure~\ref{densplot_sfr}).   Galaxies in the density regime of log($\rho_{gal}$) $\sim$ $-$1.0 show a range in measured SFRs of $\sim$4.4 dex (factor of $\sim$25,000), clearly indicating that the local density is not the sole driver for either high or low levels of star formation.   Despite the large scatter in SFR as a function of density, the binned median values again show a modest trend for lower SFRs  to occur in galaxies located at lower density regions.   Our weighted linear fit to the red points in Figure~\ref{densplot_sfr} (solid blue line) has a slope of 0.120 $\pm$ 0.045.   The implied drop in log(SFR) between log($\rho_{gal}$) = 0 and log($\rho_{gal}$) = $-$2.5 is -0.30 $\pm$ 0.14 (i.e., a drop in SFR by a factor of 2.0$^{+0.8}_{-0.6}$).   This drop in SFR as a function of local density is certainly non-negligible, but once again is of modest statistical significance.

\begin{figure}
\centering
\includegraphics[width=3.353in]{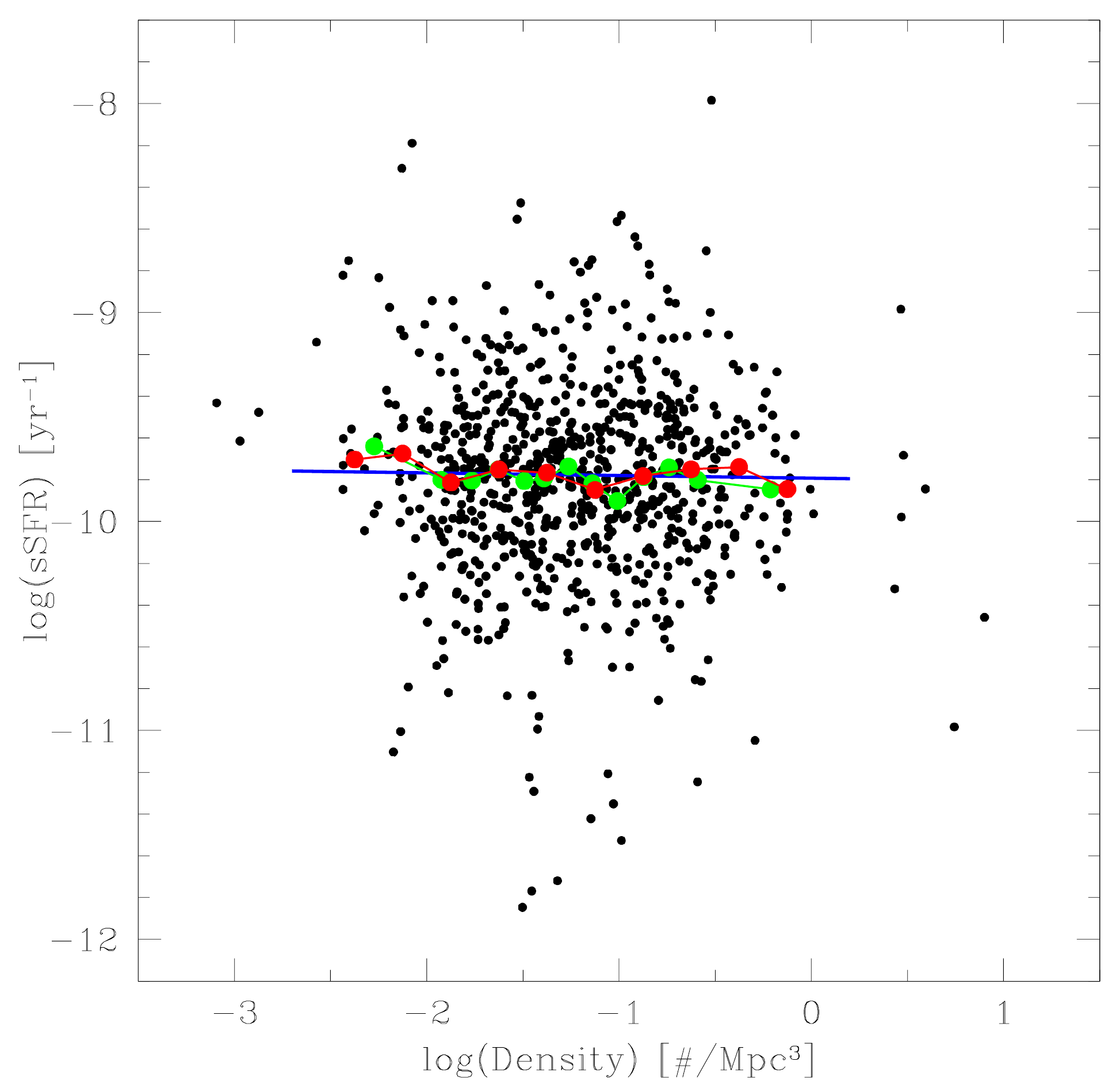}
\caption{H$\alpha$-derived specific star-formation rate (sSFR) plotted versus local density (\# Mpc$^{-3}$) for the KISS star-forming galaxies.  The total number of galaxies plotted is N=810.   The larger red and green dots represent the median sSFRs in binned ranges of density.  The green dots are binned such that there is fixed number of galaxies in each bin (N=64); the spacing between the bins varies.  The red dots represent bins of fixed width (0.25 dex) but with variable numbers of objects per bin.  Only bins with 12 or more galaxies are plotted.  Both sets of binned data show at most a weak trend for increasing sSFR as a the local density decreases.  The solid blue line represents a weighted linear fit to the red points, as described in the text.  The slope of the fitted line is consistent with zero. \label{densplot_ssfr}}
\end{figure}

For completeness we present a plot of the logarithm of the specific star-formation rate (sSFR) {\it versus} the local density in Figure~\ref{densplot_ssfr}.   The KR2 sample exhibits a large scatter in sSFR similar to what is seen in the other plots in this series.  In the case of the sSFR, however, there is little discernible trend with galaxian density.   The formal linear fit to the binned median values (red points) delivers a slope of  $-$0.013 $\pm$ 0.032, consistent with zero slope and no trend in sSFR with density.   Presumably this result arises from the fact that while log(SFR) shows a weak downward trend with decreasing density, so too does M$_*$ (Figure~\ref{densplot_mass}).   The net result is for the two effects to approximately cancel in the derivation of sSFR = SFR/M$_*$, resulting in the lack of a trend with density.

\subsection{Discussion}

It would be easy to over-interpret the results presented above.  For example, a simple-minded explanation of the modest trends seen in Figures~\ref{densplot_abun} and~\ref{densplot_sfr} could be that galaxies in lower density environments undergo fewer interactions with other galaxies, leading to lower rates of star-formation activity and lower levels of chemical enrichment.  Although this could in fact be the correct interpretation, the lack of a trend in Figure~\ref{densplot_ssfr} (sSFR vs density) seems to imply that there is essentially no dependance of the mass-normalized SFR on environment.   In this case, the decreasing median SFRs with decreasing density seen in Figure~\ref{densplot_sfr} would be due to the tendency for the galaxies located at lower densities to be of lower mass than those at higher densities (Figure~\ref{densplot_mass}).  Lower-mass galaxies will naturally have lower SFRs, on average, simply by virtue of being smaller.   This fact alone could result in the trends found in Figure~\ref{densplot_sfr}.  

An analogous argument can be made about the trend shown in Figure~\ref{densplot_abun} for the median abundances to decrease with decreasing density.   The existence of a relationship between galaxy mass and metal abundance (M-Z relation) has been known for years \citep[e.g.,][]{lequeux1979, tremonti2004}, in the sense that lower mass galaxies possess lower metallicities.   The recent study by \citet{hirschauer2018} derived a linear M-Z relation with a slope of 0.451 per dex in terms of the logarithm of the stellar mass, using precisely the same metallicities employed in the current study.  The observed drop in the value of log(M$_*$) shown in Figure~\ref{densplot_mass} is 0.38 dex.  Using the \citet{hirschauer2018} M-Z relation slope we infer that a systematic drop in the stellar mass by this amount would translate into a corresponding drop in the observed O/H abundance of -0.171 dex.   This value is remarkably close to the observed drop in abundance shown in Figure~\ref{densplot_abun} of $-$0.166 $\pm$ 0.069 dex.   In other words, the observed drop in oxygen abundance with decreasing local density can be explained entirely by the expected trends predicted by the M-Z relation.  

We conclude that our analysis of the global properties of the KR2 ELG sample as a function of environment shows modest trends of decreasing stellar mass, luminosity, metallicity, and star-formation rate with decreasing galaxian density.  However, the observed changes in abundance and SFR can be fully explained in terms of the drop in the stellar mass of galaxies in low-density environments.  Whether or not a change in the characteristic stellar mass of galaxies as a function of density is a universal phenomenon is unclear.   The evidence from the current study suggests that any environmental dependency of parameters like metallicity and SFR are essentially non-existent, with any observed changes simply tracking the lower masses of the galaxies in the lower density environments.

\section{Properties of the KISS ELGs within the Bo\"otes Void}\label{sec:void}

Our global analysis looking for possible environmental dependence on the metallicities and SFRs of the KISSR galaxies has failed to reveal any strong signatures.   Turning now to the KR2 galaxies that are actually located within the Bo\"otes Void, in this section we identify the void-dwelling population of KISS SFGs and compare their properties to a representative sample of KR2 galaxies located within higher density environments.

\subsection{Definition of Bo\"otes Void sample}

We utilize our own density analysis to define the boundaries of the Bo\"otes Void used of this study.  Historically, the void has been determined to be located at a redshift of $\sim$15,000 \kms, with a radius of $\sim$3100 \kms ($\sim$44 Mpc), and an approximate center of RA = 14$^h$ 48$^m$ and Dec = 47$^\circ$ \citep{koss83}.   Early spectroscopy of galaxies in the direction of the void revealed that it is not completely empty, but rather is threaded with low-density filaments \citep[e.g.,][]{tifft1986, moody1987, strauss1988, weistrop1988, dey1990, weistrop1992}.   Since our slice through the void is at a declination slightly south of the putative center, we do not necessarily expect the void ``center" location we find to agree precisely with previous studies.   

\begin{deluxetable*}{ccccccccccc}
\tabletypesize{\footnotesize}
\tablewidth{0pt}
\tablecaption{Properties of KISSR Galaxies located within the Bo\"otes Void\label{tab:KISSBV}}

\tablehead{
 \colhead{KISSR} &  \colhead{RA} &  \colhead{DEC} &  \colhead{d$_{cent}$} &\colhead{log($\rho_{gal}$)} &  \colhead{B$_o$} &  \colhead{(B$-$V)$_o$} &  \colhead{M$_B$} &  \colhead{log(M$_*$)} &  \colhead{log(O/H)+12} &  \colhead{log(SFR)}  \\
  &  \colhead{degrees} &  \colhead{degrees} &  Mpc & \colhead{\#/Mpc$^3$} &  &  &  &  \colhead{M$_\odot$}  &  &  \colhead{M$_\odot$/yr} \\
 (1)  & (2)  & (3)  & (4)  & (5)  & (6)  & (7)  & (8)  & (9)  & (10) & (11)
}
 
\startdata
 1654 & 206.988500 &  43.141650 & 42.1 & -2.39 & 17.46 &   0.81 & -19.28 &  10.23 &   8.96 & \ 0.68 \\
 1686 & 209.968028 &  44.255750 & 37.2 & -1.91 & 17.14 &   0.65 & -19.80 &  10.58 &   9.07 & -0.07 \\
 1688 & 210.125417 &  44.037919 & 47.8 & -2.03 & 17.84 &   0.62 & -19.28 &   \ 9.89 &   8.72 &  \ 0.37 \\
 1706 & 211.083472 &  43.874750 & 42.1 & -1.79 & 17.31 &   0.70 & -19.13 &  10.17 &   8.91 &  \ 0.47 \\
 1714 & 211.357528 &  43.504125 & 38.4 & -2.13 & 20.05 &   0.70 & -16.97 &   \ 8.03 &   7.76 & -0.28 \\
 1715 & 211.360833 &  43.471147 & 37.9 & -2.20 & 16.78 &   0.62 & -20.24 &  10.59 &   9.02 &  \ 0.91 \\
 1719 & 211.619222 &  42.636144 & 35.2 & -2.43 & 18.87 &   0.82 & -18.10 &   \ 9.58 &   8.96 & -0.03 \\
 1768 & 216.002917 &  44.232869 & 44.9 & -2.14 & 16.64 &   0.85 & -20.56 &  10.95 &   8.80 & -0.05 \\
 1780 & 216.500750 &  43.535292 & 25.9 & -2.97 & 18.20 &   0.31 & -18.80 &   \ 9.44 &   8.49 & -0.17 \\
 1781 & 216.513861 &  43.528875 & 23.4 & -3.09 & 18.02 &   0.67 & -18.94 &   \ 9.90 &   8.81 &  \ 0.46 \\
 1800 & 217.960111 &  43.087542 & 57.2 & -2.14 & 18.57 &   0.42 & -18.76 &   \ 9.54 &   8.31 & -0.27 \\
 1819 & 218.756139 &  43.945436 & 49.9 & -2.14 & 19.70 &   0.53 & -17.58 &   \ 8.93 &   8.16 & -0.16 \\
 1826 & 219.328694 &  43.238956 & 42.4 & -2.32 & 18.77 &   0.36 & -17.53 &   \ 9.06 &   8.30 & -0.98 \\
 1855 & 221.383389 &  43.241756 & 26.9 & -2.14 & 17.09 &   0.57 & -19.40 &  10.09 &   8.83 &  \ 0.32 \\
 1857 & 221.604111 &  43.832222 & 29.2 & -1.98 & 15.77 &   0.67 & -20.69 &  10.84 &   9.12 &  \ 0.86 \\
 1860 & 221.634250 &  43.655119 & 42.0 & -2.32 & 16.74 &   0.61 & -19.56 &  10.00 &   8.89 &  \ 0.26 \\
 1866 & 222.035556 &  43.094961 & 25.5 & -2.12 & 18.78 &   0.60 & -17.72 &   \ 9.21 &   8.41 & -0.30 \\
 1876 & 223.473306 &  44.013272 & 36.7 & -2.09 & 16.92 &   0.64 & -19.44 &  10.33 &   8.90 &  \ 0.38 \\
 1890 & 225.602389 &  43.865211 & 32.1 & -2.27 & 17.90 &   0.79 & -19.21 &  10.38 &   9.06 &  \ 0.41 \\
 1892 & 225.720056 &  42.678953 & 29.6 & -1.71 & 17.91 &   0.73 & -19.17 &  10.35 &   9.10 &  \ 0.38 \\
 1895 & 225.786111 &  42.648139 & 37.9 & -2.08 & 16.76 &   1.05 & -19.60 &  10.84 &   9.15 &  \ 0.58 \\
 1900 & 226.123944 &  43.173114 & 25.1 & -2.16 & 17.30 &   0.46 & -19.23 &   \ 9.75 &   8.91 &  \ 0.31 \\
 1912 & 227.127528 &  42.866850 & 44.3 & -1.90 & 17.36 &   0.62 & -18.93 &   \ 9.90 &   8.92 & -0.00 \\
 1943 & 230.726556 &  42.904364 & 27.9 & -2.26 & 20.63 &   0.87 & -16.32 &   \ 9.09 &   8.01 & -0.50 \\
 1961 & 233.076056 &  43.374992 & 29.9 & -2.25 & 16.88 &   0.76 & -19.99 &  10.52 &   9.14 &  \ 0.60 \\
 1962 & 233.085278 &  43.374986 & 29.6 & -2.40 & 18.16 &   0.57 & -18.69 &   \ 9.54 &   8.62 & -0.14 \\
 1969 & 233.377194 &  42.694956 & 44.6 & -1.59 & 19.30 &   0.97 & -17.81 &   \ 9.98 &   9.08 & -0.19 \\
 1998 & 235.227917 &  43.515356 & 45.1 & -1.69 & 18.20 &   1.00 & -18.87 &  10.25 &   8.94 & -0.10 \\
 2032 & 237.021583 &  43.120328 & 43.5 & -1.82 & 18.67 &   0.65 & -18.28 &   \ 9.71 &   8.94 & -0.35 \\
 2056 & 237.940444 &  43.457872 & 45.6 & -2.14 & 17.62 &   0.71 & -19.33 &  10.28 &   8.98 &  \ 0.28 \\
 2088 & 240.834250 &  43.057700 & 49.7 & -2.03 & 17.73 &   0.59 & -19.01 &   \ 9.90 &   8.94 & -0.44 \\
 2090 & 240.916250 &  43.144914 & 49.8 & -2.02 & 17.58 &   0.45 & -19.17 &   \ 9.80 &   8.87 & -0.50 \\
 2122 & 242.550278 &  43.317939 & 54.2 & -2.13 & 18.22 &   0.60 & -18.56 &   \ 9.70 &   8.78 &  \ 0.02 
\enddata
\end{deluxetable*}

Starting with the approximate RA, central velocity and radius found in the literature, we adjusted the position and size of the red circle shown in Figures~\ref{cone30} and~\ref{denscone}.   Most of our focus was on the comparison sample cone diagram (Figure~\ref{denscone}a).   We adjusted the circle until it maximized the inclusion of as many of the low density objects (green points) as possible while excluding as many of the intermediate density sources (blue and black points) as possible.  Our final void center was very close to the canonical values listed above.   We adopt a velocity = 15,000 \kms and RA = 14$^h$ 50$^m$.  The final radius of the void was set at 3200 \kms \ (45.7 Mpc), slightly larger than what is typically used.  Choosing a circular boundary to denote the void region was based on a combination of the simplification that this choice imparts on the analysis plus the fact the the visual appearance of the low density region in Figure~\ref{denscone}a is essentially circular.   We stress that the precise location and shape of the void boundary is not crucial for our analysis.   

Careful examination of Figure~\ref{denscone}b reveals a total of 28 KR2 galaxies located within the void.  Most are located in the outer portion of the void.   The object closest to our defined center has a velocity offset of 1637 \kms, slightly outside the midpoint between the center and outer edge of the void.   In other words, the central 50\% of the void (in terms of linear distance) is completely empty of KR2 galaxies (and is largely empty of comparison galaxies).   We modify the KR2 void sample slightly, as follows:  (i) One of the KR2 galaxies located within the void is classified as an AGN (LINER).   We remove it from the sample considered below since we are primarily interested in properties associated with star formation.  (ii)  We add six KR2 galaxies located just outside the red circle shown in Figure~\ref{denscone}b that are located in the lowest density regime (green points, log($\rho_{gal}$) $<$  $-$1.9 Mpc$^{-3}$)  The reasoning behind adding these additional objects to the KR2 void sample is that (1) they are located in density environments equivalent to the void itself, and (2) there is no reason to assume that the void is perfectly spherical.  Adding these six galaxies increases our sample of void SFGs galaxies to N = 33.   Table~\ref{tab:KISSBV} lists the individual KISSR galaxies located in the extended Bo\"otes Void along with some of their key properties.

We note in passing that despite the fact that the KR2 footprint only covers a modest fraction of the Bo\"otes Void, we still have a robust sample of star-forming galaxies with measured abundances and SFRs available to carry out our analysis.   Because of the limited overlap, most of the star-forming galaxies discovered in the void in previous studies are not found in our current KR2 sample (see \S 1).

We also create a sample of KR2 galaxies located in higher density regions of the survey field.   It is this sample of KISSR galaxies that will form the basis for the comparison of physical properties between the void and higher-density samples.  The reason for using the KR2 galaxies for this comparison is because they possess precisely the same selection function and depth as the galaxies in the void sample.  Furthermore, all KR2 galaxies have essentially identical photometric and spectroscopic data.  They represent the optimal higher density environment sample to compare against the properties of the KR2 void galaxies.  

The definition of the KR2 high density comparison sample has been done to avoid any unwanted biases.  We select all KR2 galaxies that have the following characteristics:  (i) star-forming galaxy (all AGN removed); (ii) redshifts that place them at the same distance as the void sample; (iii) located in the moderately high or high density regimes (log($\rho_{gal}$) $>$  $-$1.2 Mpc$^{-3}$).   There are a total 58 KR2 galaxies that fulfill these criteria.

\subsection{Properties of the KR2 Galaxies Located within the Bo\"otes Void}

\begin{deluxetable*}{cccccc}
\tabletypesize{\footnotesize}
\tablecaption{Comparison of Properties for KISS Galaxies: Galaxies Located in the Bo\"otes Void {\it versus} Galaxies in High Density Regions\label{tab:Props}}
\tablehead{
 \colhead{Property} &  \colhead{Sample} &  \colhead{Mean} &  \colhead{Median} &\colhead{Minimum} &  \colhead{Maximum}   \\
 (1)  & (2)  & (3)  & (4)  & (5)  & (6) 
}
\startdata
 log($\rho_{gal}$) & Bo\"otes Void & -2.14 $\pm$ 0.05 &  -2.14  &  -3.09  &  -1.59 \\
 \lbrack Mpc$^{-3}$\rbrack & High Density &  -0.93 $\pm$ 0.03 &   -0.97 &  -1.20  &   -0.24 \\
\\
redshift z & Bo\"otes Void & 0.0517 $\pm$ 0.0012   &  0.0543    &   0.0403   &    0.0647 \\
& High Density &  0.0543 $\pm$ 0.0009  &  0.0543   &    0.0401   &    0.0619 \\
\\
\hline
\\
(B$-$V)$_o$ & Bo\"otes Void & 0.67 $\pm$ 0.03   &   0.65    &  0.31   &    1.05  \\
& High Density & 0.65 $\pm$ 0.02 &   0.63   &   0.41    &   0.99   \\
\\
 M$_B$ & Bo\"otes Void & -18.91 $\pm$ 0.19 & -19.13 & -16.32 & -20.69 \\
& High Density & -18.93 $\pm$ 0.13 & -18.75 & -17.00 & -21.74  \\
\\
log(M$_*$) & Bo\"otes Void & 9.92 $\pm$ 0.11 & 9.90 & 8.03 & 10.95 \\
 \lbrack M$_\odot$\rbrack & High Density & 9.86 $\pm$ 0.08 & 9.84 & 8.72 & 11.09 \\
\\
log(O/H)+12 & Bo\"otes Void & 8.78 $\pm$ 0.06 & 8.91 & 7.76 & 9.15 \\
& High Density & 8.78 $\pm$ 0.04 & 8.87 & 7.92 & 9.16 \\
\\
log(SFR) & Bo\"otes Void & 0.08 $\pm$ 0.08 & -0.00 & -0.98 & 0.91\\
 \lbrack M$_\odot$/yr\rbrack & High Density & 0.04 $\pm$ 0.07 & -0.06 & -0.76 & 2.32 \\
\enddata
\end{deluxetable*}

\begin{figure}
\centering
\includegraphics[width=3.35in]{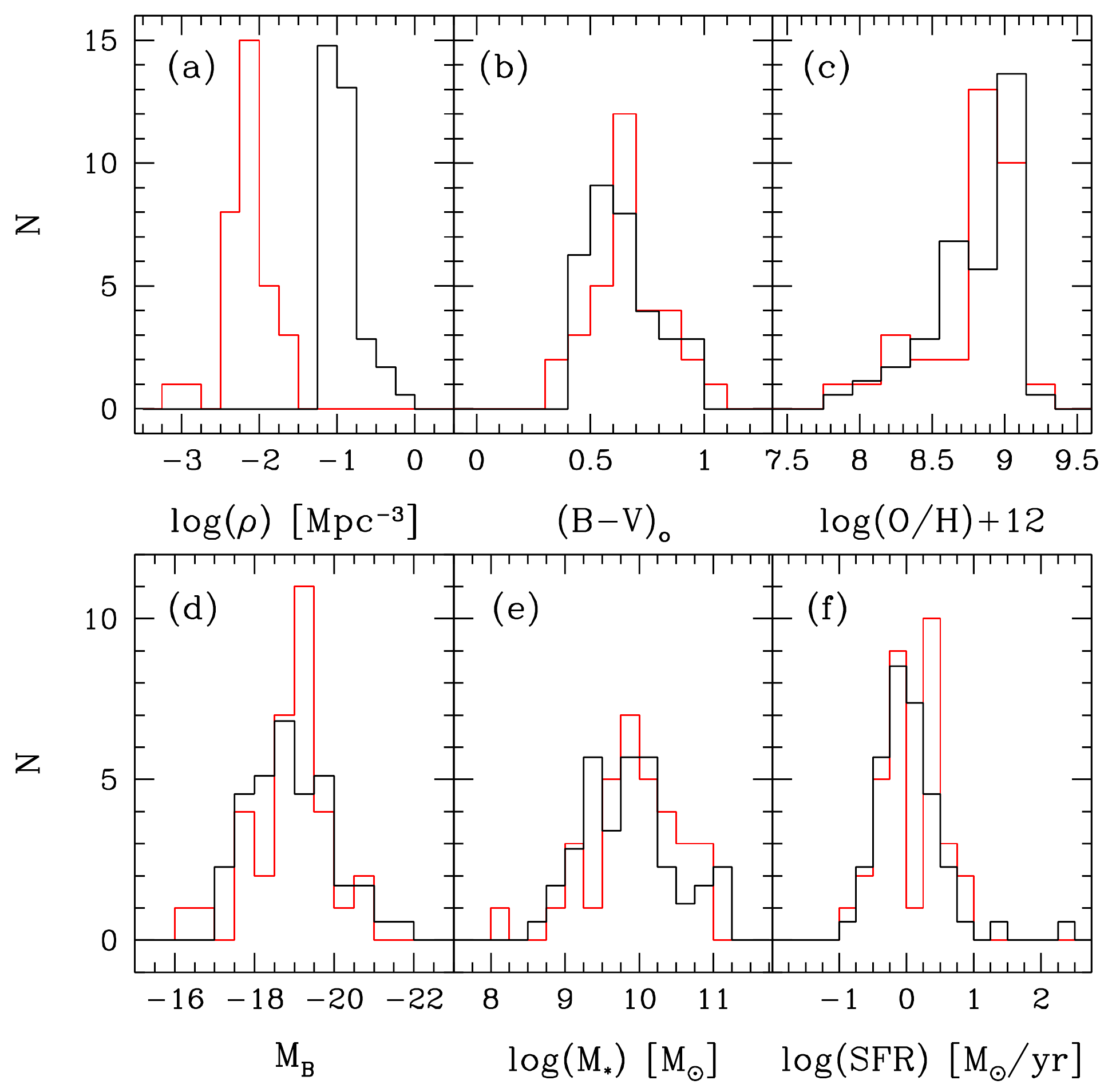}
\caption{Histograms presenting the properties of the KISSR galaxies located within the Bo\"otes Void (N=33; red histograms), as well as the same properties measured for a comparison sample of KISSR galaxies located in high density regions (N=58; black histograms).   The high density comparison sample was chosen to match as closely as possible the redshift range of the void galaxies.  The black histograms are all scaled down so that the areas under the red and black histograms are equal.  The local density histograms plotted in panel (a) show that the two samples do not overlap in terms of their environment; the mean densities differ by a factor of 16.5.   Despite this, the histograms of the other characteristics shown show strong similarities between the low and high density samples. \label{Props}}
\end{figure}

We present a direct comparison of the properties of the KR2 Bo\"otes Void galaxies against the KR2 galaxies located in higher-density environments in Table~\ref{tab:Props}  and Figure~\ref{Props}.   Figure~\ref{Props} presents histograms of observed and derived properties  for the galaxies in the two samples, while Table~\ref{tab:Props}  presents the sample mean (and its formal error) and median values for the various properties for both the void and high-density environment galaxies, as well as their minimum and maximum values.    The error in the mean is computed in the standard way ($\sigma_{property}$/$\sqrt{N}$), where\ $\sigma_{property}$ is the sample standard deviation for the property.   We note that this determination of the error in the mean is only approximate in the cases where the property in question is not well represented by a normal distribution (e.g., abundance, density), but in all cases it should serve to illustrate the degree to which the differences in the means of the properties between the high- and low-density samples are statistically significant.

The first two properties listed in Table~\ref{tab:Props} are the values of the local galaxy density and spectroscopic redshifts for the galaxies in the two samples.  As suggested by the selection criteria discussed above, the two samples have non-overlapping ranges of local density (Figure~\ref{Props}a), with the void sample having a mean density 1.22 dex (a factor of 16.5 times) lower density than the high-density KR2 galaxies.   If there are any environment-driven differences between galaxies located in the high- and low-density regions these two samples should be able to reveal them.  The mean and median redshifts of the void and comparison samples are very similar, due to the 2nd selection criterion used to define the high-density sample.  By matching the redshift distributions in this way, we effectively eliminate any distance-related biases that might affect the properties of the two samples.  Since both samples were selected by KISS with the same selection function \citep[e.g.,][]{salzer2000}, and since they have very nearly identical redshift distributions, any differences in properties present between the two samples should be due to the dramatically different environments probed by the two samples.

The mean density of the Bo\"otes Void galaxies given in Table~\ref{tab:Props} (log($\rho$) = $-$2.14 galaxies/Mpc$^{-3}$) corresponds to an under density $\delta_v$ of $-$0.91 if we adopt the median density of the SDSS+UZC galaxies from \S~\ref{sec:dens} for the global average density $\langle \rho \rangle$.   We note that some portions of the void have densities nearly a factor of 10 lower than the mean value, as evidenced by the minimum value listed in Table~\ref{tab:Props}.

Despite the strong differences in the density environments of the two subsets of KR2 galaxies, the properties of the galaxies in the void and high-density samples are very similar.    Panels (b) through (f) in Figure~\ref{Props} show the distributions of (B$-$V)$_o$ color, oxygen abundance, B-band absolute magnitude, stellar mass, and SFR for the KR2 galaxies in the Bo\"otes Void overlaid with the same parameters for the high-density sample.  No strong differences are seen in the overall distributions of any of these properties between the high- and low-density galaxies, although the histograms do show numerous small-scale differences.   For example, there is a sharp peak in the absolute B-band magnitudes for the KR2 void galaxies at $\sim-$19.0 (panel d) even though the overall distributions for the void and high-density luminosities are quite similar.   Likewise, the log(SFR) distributions show good overall agreement but exhibit strong differences in a few of the histogram bins between log(SFR) = 0.0 and 0.5..

K-S tests applied to the CDFs for the data shown in panels (b) through (f) of Figure~\ref{Props} return results that in most cases favor the hypothesis that the void and high-density distributions are drawn from the same parent populations.   For example, the K-S test applied to the abundance histograms (panel c) finds that there is an 90.4\% probability that the two distributions of log(O/H)+12 values are drawn from the same parent population of metallicities, while for the absolute magnitudes (panel d) there is a 92.4\% chance of this being the case.    The corresponding probabilities for the other three distributions are 76.3\% for log(M$_*$), 63.5\% for the (B$-$V)$_o$ color, and 39.5\% for the log(SFR) values being drawn from the same parent population of objects.  Most of the differences between the void and high-density distributions appear to be small-scale sampling issues, to which the K-S test is particularly vulnerable.   Only the log(SFR) K-S test returns a probability of less than 60\% that the two distributions come from the same population of objects.   The smaller probability is the result of the two discrepant bins pointed out above.   We conclude that there is no strong evidence to support the hypothesis that the void and high-density samples are drawn from different parent populations.  In fact, the evidence tends to support just the opposite conclusion: the properties of the KR2 galaxies located in the Bo\"otes Void are quite similar to those found in the high-density regions nearby.

In further support of the similarities between the void and high-density galaxies, the mean and median values of these properties listed in Table~\ref{tab:Props} show essentially no differences between the two samples.   In particular, the mean value of log(O/H)+12 is identical for the void-dwelling galaxies and the objects found at the higher densities, while the mean SFR differs by only 0.04 dex ($\sim$10\%), with the high-density galaxies exhibiting a slightly lower mean value.

Similar to our results from the previous section, we find no strong differences in the properties of the KR2 galaxies located within the Bo\"otes Void and those found in higher density regions.   The mean and median values for each of the parameters considered in our analysis are essentially the same for the galaxies in the two density regimes.   While the minimum values shown in Table~\ref{tab:Props} reveal that there are a few void galaxies with extreme properties (lower luminosities, masses, abundances and SFRs),  the overall distributions are indistinguishable.  We conclude that the KR2 galaxies found in the Bo\"otes Void have evolved in a manner that is quite similar to the star-forming galaxies located at much higher densities.

\section{Discussion and Implications}

Our primary result is that there are no significant differences in the metallicities or the star-formation rates between the KR2 galaxies located in higher density regions and those found in voids.   This lack of environmental dependence is found in both our global study (\S~\ref{sec:global}) as well as our detailed investigation of the KR2 galaxies within the Bo\"otes Void (\S~\ref{sec:void}).  These findings are consistent with most previous studies, after accounting for sample selection differences.

Several previous studies have considered the metallicities of void galaxies \citep[e.g.,][]{peimbert1992, cruzen2002, kreckel2015, douglas2018}, with the latter two studies focusing primarily on dwarf galaxies.   In all cases, the authors conclude that there are no major differences when comparing metal abundances between galaxies in voids and those located in denser regions.   These results are completely consistent with our findings.   These previous studies tended to be limited to modest-sized samples with less-than-ideal control samples (or none at all).  One of the obvious strengths of the current study is that 100\% of our sample of SFGs possess metallicity estimates, allowing us to construct unbiased samples of high- and low-density galaxies.

Slightly greater care must be exercised when considering the results of previous studies that have looked into the star-formation properties of void {\it versus} non-void galaxies.   Many studies have found that galaxies in voids show systematically higher equivalent width emission lines \citep[e.g.,][]{grogin2000} and higher SFRs and sSFRs \citep[e.g.,][]{rojas2005, ricciardelli2014, moorman2016} when compared to galaxies in higher density regions.   These findings reflect real differences between the galaxy populations found in voids and those found in higher density environments.  

However, when exploring whether or not the local environment has an impact on the star-formation properties of galaxies, one must carefully control for these population differences.   In studies where the properties of the non-void comparison sample were matched to those of the void sample (e.g., similar colors, masses, gas content), quantities like sSFR were found to be similar across environments \citep[e.g.,][]{patiri2006, ricciardelli2014, moorman2016}.   The results of these latter studies are in good agreement with our findings.   We note that we use the KR2 sample in this study both as the source of the properties of the void galaxies as well as for the higher-density comparison sample.   Hence, our sample is property-matched from the start.  This is one of its advantages, and is why it is sensitive to any possible environmental effects on the constituent galaxies.  

It is worth noting that our density analysis measures the local galaxy environment on scales of hundreds of kiloparsecs to a few megaparsecs.   We have relatively little spatial information on scales of tens of kiloparsecs, where galaxy-galaxy interactions would be taking place.   Hence, we stress that in our study the local ``environment" is being measured on scales that may not capture the full picture of what is driving the star-formation activity.  Of course, the KR2 sample should correctly reflect {\it all} star-formation activity, regardless of the nature of the triggering event.  Furthermore, while this caveat may be an issue when relating environment to SFR, it should not matter when comparing environment to metallicity.   This is because the latter parameter is built up over the lifetime of the galaxy and is less likely to be affected significantly by individual, small-spatial-scale interactions.

Our study appears to rule out any significant differences in the metallicities and SFRs of galaxies as a function of local environment.   It is important to remember, however, that the samples used in this study are not sensitive to the lowest-mass star-forming galaxies at the distance of the Bo\"otes Void.  While the KR2 galaxies in the Void and high-density samples are well-represented down to masses of 10$^{9.25}$ M$_\odot$ and to luminosities of M$_B$ $\sim$ $-$18.0 (i.e., comparable to the mass and luminosity of the LMC), the numbers for truly low-mass / low-luminosity systems drops off rapidly below these values.  Only a handful of KR2 galaxies qualify as true dwarfs at these distances.    Hence, our study cannot rule out the possibility that some significant amount of environmental dependence on metallicity and/or SFR/sSFR is only manifested in very low-mass dwarf galaxies with masses and luminosities much less than the SMC (i.e., M$_B$ $\gtrsim$ $-$16.5).  The recent studies of \citet{pustilnik2016} and \citet{kniasev2018} propose that O/H values of low-luminosity dwarfs found in nearby voids are systematically lower than those found in higher density regions.   We find these studies tantalizing.   We note, however, that the \citet{kreckel2015} study which found no differences in the metallicities of void and field samples also focused on low-luminosity galaxies (M$_B$ = $-$12.0 -- $-$15.9).

\section{Summary and Conclusions}

We have carried out a detailed analysis of the density distribution of galaxies located in the direction of the Bo\"otes Void in order to probe for any potential environmental impacts on the metallicities and star-formation rates (SFRs) of galaxies.  Our sample of star-forming galaxies is the second H$\alpha$-selected catalog of the KPNO International Spectroscopic Survey \citep[KISS;][]{salzer2000, gronwall2004b}, which consists of a narrow survey strip that crosses through the void.  All 820 star-forming galaxies in our sample possess self-consistent metallicity and SFR estimates.  We also utilize a deep comparison sample consisting of two complementary redshift surveys \citep{falco1999, sdss7} to map out the distribution of galaxies in our field.  A total of 14,577 galaxies are used to define the density distribution for this study, with the derived density values covering over four orders of magnitude.

We adopt two approaches to assess the degree to which the environment impacts the properties of galaxies.   In the first case we explore the dependence of galaxian properties with local density for the entire volume covered by the second H$\alpha$-selected KISS (KR2) catalog.  In the second we identify the KR2 star-forming galaxies located within the Bo\"otes Void and compare directly their properties with a matched comparison sample of KR2 galaxies located in nearby high-density regions.

Our global analysis looks at the distribution of several key parameters as a function of local density.  These include stellar mass, B-band luminosity, oxygen abundance, SFR and specific SFR.   In all cases, we find that the parameters cover a large range at all densities: there are high-mass and low-mass galaxies at both high and low densities, and there are both low-metallicity and high-metallicity galaxies at the full range of densities probed by our project.   Nonetheless, the mean values of these properties do show weak but discernible trends with density.  As seen in Figures~\ref{densplot_mass} through~\ref{densplot_sfr}, the stellar mass, M$_B$, metallicity, and SFR all show decreasing trends with decreasing local density.   While this would appear to be direct evidence for environmental impact on both the chemical evolution and star-formation properties of the KR2 galaxies, we interpret the trends differently.   The change in metallicity with density follows precisely the trend one would predict from the mass-metallicity relation of galaxies and the observed downward trend of galaxy mass seen in Figure~\ref{densplot_mass}.  Hence, we conclude that there is no real impact of environment on the chemical evolution of the galaxies in the KR2 sample.   Likewise, the downward trend of SFR with density seen in Figure~\ref{densplot_sfr} is due entirely to the observed mass-density trend.  When we plot the mass-normalized specific SFR {\it versus} density (Figure~\ref{densplot_ssfr}) no density-dependent trend is seen.

For the analysis of the KR2 galaxies that actually reside in the Bo\"otes Void, we first analyzed the density distribution as defined by our deep comparison redshift survey data in order to properly define the location of the void boundaries.  This was necessary because the KR2 sample covers only a limited section of the overall void.   Once our void sample of KR2 star-forming galaxies was defined (N=33), we constructed a matched comparison sample of KR2 galaxies located within the high-density regions flanking the void.   This sample (N=58) was chosen carefully to match the redshift range covered by the void galaxies, in order to ensure that no biases were present.  A direct comparison between the properties of the void and high-density samples revealed no significant differences between the two samples, despite the fact that the high-density galaxies are located in regions with a mean local density 16.5 times higher than the average density of the Bo\"otes Void sample.

In summary, we find no evidence that the properties of galaxies depend strongly on the local environment.  While we do find a weak trend of decreasing stellar mass (and a corresponding decrease in B-band absolute magnitude) with decreasing density, we do not find any indication that the metallicity of galaxies or their SFRs depend on environment.   It is the case that our current sample does not probe down to the lowest masses and metallicities at the distance of the Bo\"otes Void, so we cannot rule out the possibility that environmental effects may impact the lowest mass galaxies.  For masses above $\sim$10$^{9.0-9.5}$ M$_\odot$, however, we believe our study places strong limits on the degree to which properties like metal abundance and SFR are impacted by the environment within which a galaxy is located.   This in turn implies that the processes of chemical evolution and star formation  are largely governed by internal processes, at least for non-merger systems.   Since the metallicity of a galaxy is a measure of its time-integrated star formation history, our results would seem to imply that the assembly history of galaxies that are {\it currently} located in vastly different density environments may well have been similar.   This might result naturally if most of the mass assembly happens at early times before the large differences in local galaxy density seen in the modern universe developed.

\acknowledgements

We acknowledge the significant contributions to this project with the many collaborators who have contributed to the KISS project over the years.  In particular, many thanks to Caryl Gronwall, Todd Boroson, Steven Janowiecki, Jessica Werk, Vicki Sarajedini, Anna Jangren, Janice Lee, Jason Melbourne, Laura Chomiuk,  Anna Williams, Lisa Frattare, Jos\'e Herrero, Trinh Thuan, Yuri Izotov, Alexei Kniazev and J. Ward Moody.  We acknowledge financial support for the current project from both Dartmouth College and Indiana University.  We thank the observatory support staff at the MDM Observatory for their expert assistance during the observing runs used to acquire many of the spectra used for this project.

\facility{Hiltner}

\end{document}